\documentclass[11pt]{article}
\usepackage{jheppub,amsmath, amsthm, amssymb,slashed,url}
\usepackage{graphicx}
\usepackage{epstopdf}
\usepackage{tikz}
\usepackage{xcolor}

\usepackage{etoolbox}
    \makeatletter
    \patchcmd{\maketitle}{\@fpheader}{}{}{}
    \makeatother

\usepackage{bm}
\usepackage{amssymb,amsfonts}
    
\DeclareMathOperator{\sgn}{sgn}

\def\phimax{\phi_{\scriptscriptstyle\rm max}}
\def\thetamax{\theta_{\scriptscriptstyle\rm max}}
\def\pmax{p_{\scriptscriptstyle\rm max}}
\def\sigmax{\sigma_{\scriptscriptstyle\rm MAX}}

\def\oneht{\textstyle{1\over 2} }

\def\onefourth{\textstyle{1\over 4} }

\def\OMIT#1{{}}

\def\si{^1 \hskip -0.03in S _0}
\def\siii{^3 \hskip -0.025in S _1}
\def\diii{^3 \hskip -0.03in D _1}

\newcommand{\beq}{\begin{equation}}
\newcommand{\eeq}{\end{equation}}
\newcommand{\beqa}{\begin{eqnarray}}
\newcommand{\eeqa}{\end{eqnarray}}

\newcommand{\nn}{\nonumber}

\newcommand*\circled[1]{\tikz[baseline=(char.base)]{
   \node[shape=circle,draw,inner sep=1pt] (char) {#1};}}

\begin{document}

\dedicated{NT@UW-20-08, IQuS@UW-20-01}
\title{Geometry and entanglement in the scattering matrix}
\vskip 0.5cm
\author{\bf Silas R.~Beane}
\author{\bf and Roland C.~Farrell}
\affiliation{InQubator for Quantum Simulation (IQuS), Department of Physics,\\
  University of Washington, Seattle, WA 98195.}


\vphantom{} \vskip 1.4cm \abstract{A formulation of nucleon-nucleon
  scattering is developed in which the $S$-matrix, rather than an
  effective-field theory (EFT) action, is the fundamental
  object. Spacetime plays no role in this description: the $S$-matrix
  is a trajectory that moves between RG fixed points in a compact
  theory space defined by unitarity. This theory space has a natural
  operator definition, and a geometric embedding of the unitarity
  constraints in four-dimensional Euclidean space yields a flat torus,
  which serves as the stage on which the $S$-matrix
  propagates. Trajectories with vanishing entanglement are special
  geodesics between RG fixed points on the flat torus, while
  entanglement is driven by an external potential.  The system of
  equations describing $S$-matrix trajectories is in general
  complicated, however the very-low-energy $S$-matrix ---that appears
  at leading-order in the EFT description--- possesses a UV/IR
  conformal invariance which renders the system of equations
  integrable, and completely determines the potential. In this
  geometric viewpoint, inelasticity is in correspondence with the
  radius of a three-dimensional hyperbolic space whose two-dimensional
  boundary is the flat torus. This space has a singularity at
  vanishing radius, corresponding to maximal violation of unitarity.
  The trajectory on the flat torus boundary can be explicitly
  constructed from a bulk trajectory with a quantifiable error,
  providing a simple example of a holographic quantum error correcting
  code.}

\maketitle



\section{Introduction}

\noindent In a generic quantum-mechanical scattering process, it is
usual to view an effective field theory (EFT) action as the
fundamental object from which all information about the scattering
process can be obtained. General physical principles as well as the
internal and spacetime symmetries of the system in question are
conveniently encoded in the EFT~\cite{Weinberg:1978kz}, which is given
a precise definition as a perturbative expansion in local operators
about a fixed point of the renormalization group
(RG)~\cite{Wilson:1973jj}. This view of scattering is remarkably
successful and intuitive. As scattering processes take place in
spacetime and the fundamental experimental knobs are momentum and
energy, it stands to reason that the suitable stage for describing the
process is spacetime itself. On the other hand, there are many things
going on in the scattering process which a description based on
local interactions in spacetime may not capture. For instance, the system could be highly entangled in all its degrees of
freedom, and quantum entanglement is intrinsically non-local in
spacetime.
 This raises the question of whether important effects may
be missing in field-theoretic descriptions of physical systems where
entanglement is expected to play a role. Recent work, based on earlier work in Refs.~\cite{Audenaert:2002xfl,Marcovitch:2008sxc,Calabrese:2012ew}, has investigated
this issue in free scalar field theory in $1+1$ dimensions and found evidence
for an exponential suppression of quantum correlations which can lead
to the presence of a new relative length scale associated with
geometric entanglement~\cite{Klco:2020rga}. 

Abandoning a description in terms of local operators suggests
abandoning spacetime as the stage on which the scattering is defined
and described.\footnote{For recent efforts in formulating a spacetime-independent
theory of
perturbative gauge theory scattering amplitudes, see
Refs.~\cite{Arkani-Hamed:2020gyp,Arkani-Hamed:2019vag,
Arkani-Hamed:2018bjr}.} What then is the appropriate stage on which to describe
the scattering process which renders the quantum-entangling properties
manifest?  This is the main question addressed in this paper.
Consider abandoning locality in spacetime and focusing on a more
fundamental principle: unitarity. This in turn suggests that the
unitary $S$-matrix replaces the action as the fundamental object which
characterizes the scattering process. If spacetime is the stage on
which EFT interactions are defined, what is the stage on which the
unitary $S$-matrix propagates? In this paper, it is shown that the
$S$-matrix is naturally viewed as a trajectory in a theory space that
is defined by all possible unitary trajectories. A rather surprising
and essential observation is that in special ---yet relevant--- cases,
the $S$-matrix possesses symmetry that is not visible in the EFT
description.  In particular, in these cases, the $S$-matrix possesses
a UV/IR conformal invariance which plays a critical role in all that
follows.

While the considerations in this paper are quite general, the focus
will be on the paradigmatic problem of s-wave nucleon-nucleon
scattering. This system is ideal for addressing these questions as it
constitutes the simplest kind of quantum-mechanical scattering with a
finite-range potential, it has non-trivial spin and isospin and
interesting properties as regards entanglement. Moreover the system
exhibits a non-trivial fixed-point structure which defines the EFT of
nucleon-nucleon scattering to be a perturbative expansion about a
fixed point of the
RG~\cite{Kaplan:1998tg,Kaplan:1998we,vanKolck:1998bw,Birse:1998dk}.
The EFT in question is, in some sense, the simplest low-energy EFT of
the Standard Model. At momentum transfers small as compared to the
mass of the pion, the leading-order (LO)
effective Lagrangian, constrained by spin, isospin and Galilean
invariance, can be put in the form\cite{Weinberg:1990rz}
\begin{equation}
{\cal L}_{\rm LO}
=
-\frac{1}{2} C_S (N^\dagger N)^2
-\frac{1}{2} C_T \left(N^\dagger\hat{\bm \sigma}N\right)\cdot \left(N^\dagger\hat{\bm \sigma}N\right) \ ,
\label{eq:interaction}
\end{equation}
where $N$ represents both spin states of the proton and neutron fields
and $\hat {\bm \sigma}^\alpha$ are the Pauli spin matrices. These
interactions can be re-expressed as contact interactions in the $\si$
and $\siii$ channels with couplings $C_0 = ( C_S-3 C_T) $ and $C_1 =
(C_S+C_T)$ respectively, where the two couplings are fit to reproduce
the $\si$ and $\siii$ scattering lengths. Higher-order operators
involve derivatives acting on the nucleon fields and give rise to
effective range and shape-parameter corrections. The interactions of
Eq.~(\ref{eq:interaction}) are highly singular. Defining the couplings
in dimensional regularization with the PDS scheme~\cite{Kaplan:1998tg,Kaplan:1998we}, and
choosing the renormalization scale to be the pion mass, one finds
$C_T^{\rm PDS}/C_S^{\rm PDS}=0.0824$.  This well-known suppression of
the spin-entangling operator that appears at LO in the EFT expansion
has motivated the study of entanglement in (hyper)nuclear
systems~\cite{Beane:2018oxh}, as a measure of the degree
of entanglement of interaction operators in the EFT would appear to be
required.

Forget now about the EFT description and consider a direct
construction of the nucleon-nucleon $S$-matrix. At low energies,
neutrons and protons, with two spin degrees of freedom each, scatter
via the phase shifts $\delta_{0,1}$, in the $\si$ and $\siii$
channels, respectively, with projections onto higher angular momentum
states suppressed by powers of the nucleon momenta.  Neglecting the
small tensor-force-induced mixing of the $\siii$ channel with the
$\diii$ channel, the $S$-matrix for nucleon-nucleon scattering below
inelastic threshold can be decomposed as
\begin{eqnarray}
  \hat {\bf S}(p)
  & = &
{1\over 4}\left( 3 e^{i 2 \delta_1(p)} + e^{i 2 \delta_0(p)} \right)
 \hat   {\bf 1}
\ +\
{1\over 4}\left( e^{i 2 \delta_1(p)} - e^{i 2 \delta_0(p)} \right)
  \hat  {\bm \sigma} \cdot   \hat  {\bm \sigma} \ ,
  \label{eq:Sdef}
\end{eqnarray}
where in the direct-product space of the nucleon spins,
\begin{eqnarray}
 \hat {\bf 1} \equiv \hat {\cal I}_2\otimes  \hat {\cal I}_2 \ \ \ , \ \
  \hat {\bm \sigma} \cdot \hat {\bm \sigma} \equiv \sum\limits_{\alpha=1}^3 \
\hat{\bm \sigma}^\alpha \otimes \hat{\bm \sigma}^\alpha \ ,
  \label{eq:Hil}
\end{eqnarray}
with $\hat {\cal I}_2$ the $2\times 2$ unit matrix.  It is important to
stress that this decomposition of the $S$-matrix follows from
unitarity, the symmetries of the system, and from the fact that the
nucleons are fermions. The standard procedure for obtaining the phase
shifts is to compute the scattering amplitude using the EFT and then
match onto the $S$-matrix. In this paper, a new method will be
developed for obtaining the phase shifts, which does not rely on a
spacetime picture involving local operators.  Several issues
immediately present themselves and will be discussed in turn.

First, the $S$-matrix in Eq.~(\ref{eq:Sdef}) is expressed as a
function of the momentum variable $p$ (which we take throughout as the
center-of-mass momentum). The momentum variable arises naturally in
the spacetime picture and the dependence of the phase shifts on
momentum is determined by the expansion in local operators which is
constrained by Galilean invariance. In a spacetime-independent
description of scattering, the choice of the momentum as the relevant
variable is necessarily arbitrary, and indeed it will turn out that
the choice of variable is related to the choice of parameterization of
trajectories on a Riemannian manifold.

A second issue is that of the fixed-point structure of the EFT and how
that manifests itself at the level of the $S$-matrix.  In
non-relativistic EFT, scale invariance at a fixed point is realized as
Schr\"odinger symmetry of the EFT action.\footnote{For the present
  context, see Refs.~\cite{Mehen:1999qs,Mehen:1999nd,ubi2017}.} Aside from the free theory,
which clearly admits Schr\"odinger symmetry, non-relativistic EFT
admits a non-trivial fixed point at the strongest possible coupling
consistent with unitarity (known as ``unitarity''). In the language of
the effective range expansion (ERE), this corresponds to taking the
s-wave scattering length to infinity while higher-order effective
range and shape parameters are taken to zero.  As the relation between
the scattering length and the EFT operator couplings is
renormalization scheme dependent, the EFT couplings in themselves
carry no physical, scheme-independent, information; indeed in a
mass-independent scheme the couplings are driven to infinity at
unitarity, whereas in a mass-dependent scheme (like the PDS scheme)
the couplings can be made to approach a number of order unity in units
of the characteristic physical length scales.  By contrast, as will be
seen in detail below, the $S$-matrix provides a physical
regularization of the infinite coupling limit as the fixed points
arise at finite values of the $S$-matrix. The notion of RG flow
in the EFT couplings will translate to $S$-matrix momentum flow,
which will be described by equations that exhibit the fixed point
structure of the RG.

A third issue regards measures of quantum entanglement.  A common
measure of the quantum entanglement in the EFT is the state-dependent
entanglement entropy, which is generally highly singular in quantum
field theory, much like the renormalized couplings.  One may worry
that by working directly with the $S$-matrix, one loses valuable
information regarding the nature of the entangling properties of the
interaction. However, a physical measure of the entanglement of
interaction --the entanglement power (EP)-- has recently been
developed and applied to the unitary $S$-matrix.\footnote{For previous work on the entanglement generated in non-relativistic scattering see Ref.~\cite{HARSHMAN_2005}.} The
EP~\cite{PhysRevA.63.040304,mahdavi2011cross} is a state-independent
and physical measure of the entanglement induced by interaction, for
instance by particular operators in an EFT. The EP of the
nucleon-nucleon $S$-matrix was recently considered in
Ref.~\cite{Beane:2018oxh}.

Tying these last two issues together, this paper will show that the
fixed-point structure of the $S$-matrix has geometrical significance
whose origin is closely tied with quantum entanglement. In particular,
one may view the RG fixed points as vertices on a compact
two-dimensional surface, with the $S$-matrix a trajectory that
moves on this manifold between the fixed points. In the absence of
EP the surface appears as a lattice of fixed points;
the $S$-matrix is constrained to one dimensional motion along a
geodesic, providing links between the RG fixed points. Only in the
presence of entanglement does the surface appear as a flat torus, a flat, 
compact Riemannian manifold.

In one of the more striking consequences of the geometric formulation
of scattering, inelasticity --a non-local effect in the effective
field theory--- is, in its simplest realization, in correspondence
with the radius of a three-dimensional hyperbolic space whose
two-dimensional boundary is the flat torus. This space has a
singularity at vanishing radius, corresponding to maximal violation of
unitarity.  The boundary geodesic can be explicitly constructed from a
bulk geodesic with a quantifiable error, providing a simple example of
holographic duality and quantum error correction.

This paper is organized as follows.  Section~\ref{sec:smatse} sets up
the $S$-matrix framework. First, some essential properties of the
$S$-matrix are introduced in the context of single-channel s-wave
scattering. Then various analogous properties of the nucleon-nucleon
$S$-matrix are considered, including the fixed points of the RG, the
parameterization of the $S$-matrix, the EP and the momentum flow. This
is followed by the consideration of the two specific $S$-matrices that
will be treated. The main example that will be focused on is LO in the
ERE which corresponds to the scattering length approximation (LO in
the EFT). A second example, the conformal range model, includes
effective range corrections but in a manner that is correlated between
the two channels to maintain a UV/IR conformal symmetry.
Section~\ref{sec:geosmatse} develops the geometric picture of the
$S$-matrix. First, a coordinate-independent operator-defined metric is
introduced using the notion of the Hilbert-Schmidt distance between
operators. This is followed by construction of the embedding of the
$S$-matrix unitarity constraints in the four-dimensional Euclidean
space $\mathbb{R}^4$. This embedding gives rise to the flat torus
which is the two-dimensional manifold on which the $S$-matrix
propagates. The isometries of the flat torus are then considered in
detail.  Section~\ref{sec:smattheory} is in some sense the main
section of the paper. First, the general action principle which
describes the $S$-matrix trajectories is considered. In turn, the
geodesics on the flat torus are constructed, followed by a recovery of
LO in the ERE by positing an external entangling potential. The
conformal range model is also analyzed and its potential is
found. Section~\ref{sec:inelatandholo} relaxes the unitarity condition
to consider inelastic loss. The geodesics on the resulting hyperbolic
geometry are constructed, and, in the presence of a modified external
potential, LO in the ERE is recovered up to a quantifiable error.  In
Section~\ref{sec:projs}, the geometry of the projections of the flat
torus on two- and three-dimensional spaces is considered, and it is
explicitly shown how non-conservative forces are necessary to describe
trajectories on the projected spaces. Finally,
Section~\ref{sec:summandconc} is a summary of the geometrical
$S$-matrix construction and a discussion of open questions and future
work.

\section{$S$-matrix and spin entanglement}
\label{sec:smatse}

\subsection{Phase shifts, fixed points and momentum flow}

\subsubsection*{Single-channel scattering}

\begin{figure}[!ht]
\centering
\includegraphics[width = 0.4\textwidth]{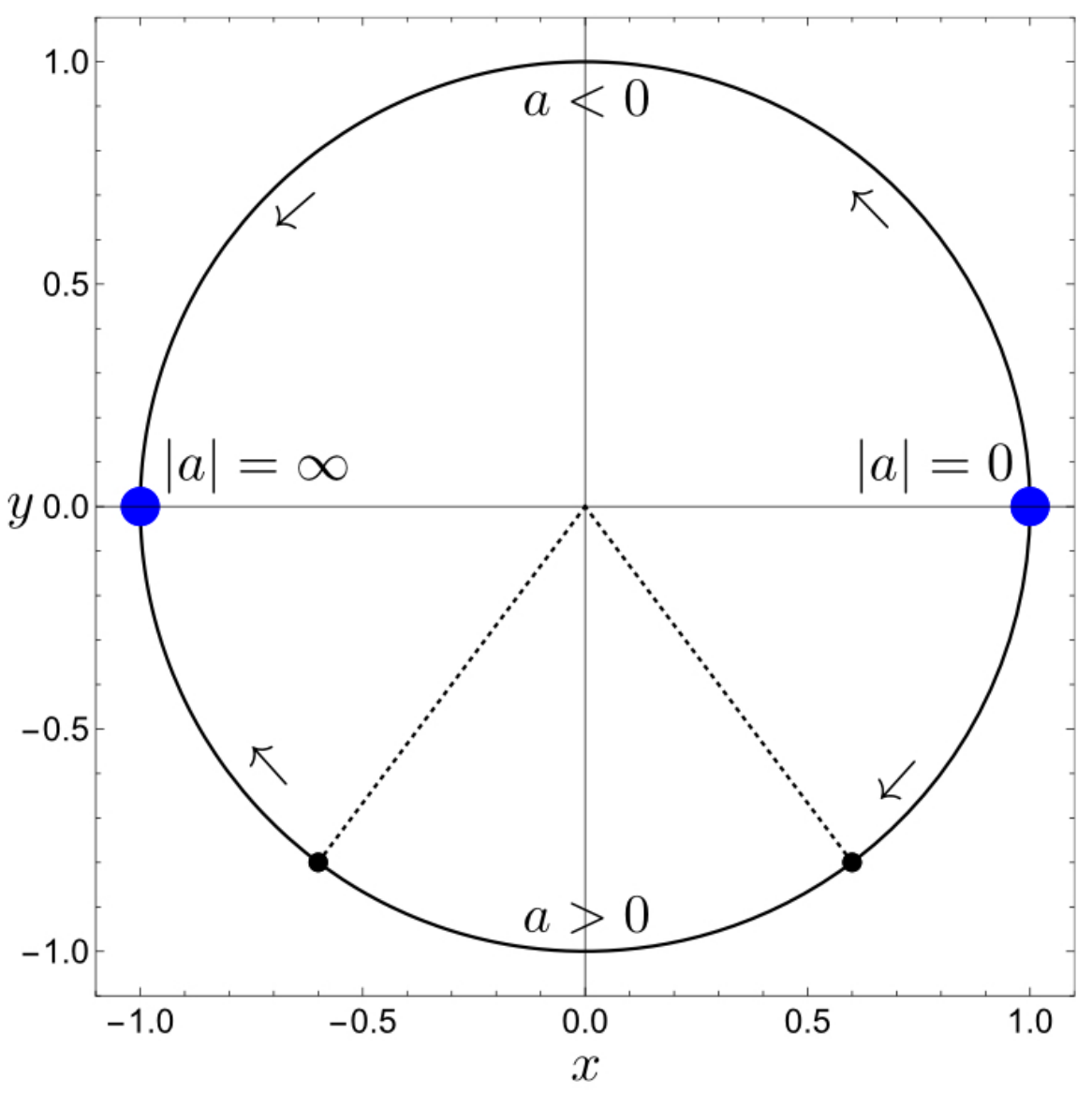}
\caption{The unitarity unit circle for single-channel scattering. In
  these coordinates, the $S$-matrix is a trajectory from the trivial
  fixed point (blue dot on the right) to the non-trivial fixed point
  (blue dot on the left). At LO in the ERE, there is a conformal
  transformation on the momenta which takes the phase shift to minus
  itself modulo $\pi/2$. The two black dots are related by a conformal
  transformation. The upper (lower) half-circle corresponds to
  negative (positive) scattering length.}
    \label{fig:SMoneD}
\end{figure}

\noindent In single-channel, non-relativistic, s-wave scattering below inelastic threshold,
the $S$-matrix in manifestly unitary form can be expressed as  
\begin{eqnarray}
  S(p) & = & e^{i 2 \delta(p)} \ =\  \frac{1+i \tan\delta(p)}{1-i\tan\delta(p)} \ ,
      \label{eq:Ssingledef}
\end{eqnarray}
where $\delta(p)$ is the phase shift and $p$ is the center-of-mass
momentum. The (scale-invariant) RG fixed points of the $S$-matrix occur
when there is no interaction, $\tan\delta(p)=0$ ($\delta=0$), or when
the interaction is maximal, $\tan\delta(p)=\infty$
($\delta=\frac{\pi}{2}$), corresponding to $S(p)= 1$ or $-1$,
respectively. The assumption that $p\cot\delta(p)$ is analytic in
$p^2$ is sufficient to give the ERE, where at LO in the ERE,
$\tan\delta(p)=-pa+{\mathcal O}(p^3)$, with $a$ the scattering
length. In the limit that $a\rightarrow 0$ ($\pm \infty$) while
effective range and shape parameters are taken to zero,
$S(p)\rightarrow 1$ ($-1$). Parameterizing the $S$-matrix as
$S=x+i y$ constrains the $S$-matrix to the unit circle
in the $x$-$y$ plane as shown in Fig.~(\ref{fig:SMoneD}).

Consider LO in the ERE and assume that all inelastic thresholds are pushed
to infinity so that $p$ is defined on the interval $[0,\infty)$. The real coordinates are
\begin{eqnarray}
x(p) & =& \frac{1- a^2 p^2}{1+ a^2 p^2} \ \ \ , \ \ \
y(p) \ =\  -\frac{2 a p}{1+ a^2 p^2}  \ .
   \label{eq:Sscefre}
\end{eqnarray}
Note that a rescaling of the momenta is compensated by a rescaling of
the scattering length; i.e. $x$ and $y$ are invariant with respect to
the dilatation $p\mapsto e^\epsilon p$, $a\mapsto e^{-\epsilon}
a$. This scale invariance implies that the fixed points of the RG are
accessed by varying the scattering lengths at fixed momenta or varying
the momenta at fixed scattering length. Hence, under the action of a
momentum dilatation $p \mapsto e^\epsilon p$, for small $\epsilon$,
the $S$-matrix beta-function can be defined:
\begin{eqnarray}
{\beta({{S}})} \equiv p\frac{d}{dp} {{S}}(p) & = & \frac{1}{\epsilon}\left( {{S}}(e^\epsilon p)- {{S}}(p) \right) \ ,
\end{eqnarray}
where ${\beta({ { S}})}$ depends on details of the dynamics and
vanishes at the fixed points where ${{S}}^2=1$. At LO in the ERE,
the $S$-matrix momentum flow is governed by
\begin{eqnarray}
p\frac{d}{dp} {{S}}(p) & = & \oneht\left( S^2-1 \right)\ .
\end{eqnarray}
For $a$ finite and fixed, the $S$-matrix trajectory begins at the trivial fixed point,
$(x,y)=(1,0)$ at $p=0$, and flows to the non-trivial fixed point at
$(-1,0)$ when $p\rightarrow\infty$. This behavior of moving along a semi-circle holds for
the full phase shift $\delta(p)$ as unitarity constrains the
$S$-matrix to one dimension and therefore there is no freedom to move
off the unit circle in the absence of inelastic effects. Importantly,
LO in the ERE has a special property beyond the scale transformation
discussed above.  Under the momentum inversion,
\begin{eqnarray}
p\mapsto \frac{1}{a^2 p} \ ,
   \label{eq:moebius1}
\end{eqnarray}
the coordinates transform as $(x,y)\to (-x,y)$ and the phase shift
transforms as $\delta(p)\rightarrow -\delta(p)\pm \pi/2$. This
symmetry will be referred to as a conformal transformation as it
leaves the phase shift invariant modulo fixed phases. However, note
that this is a symmetry transformation that interchanges the UV and
the IR. As such, this symmetry is generally broken by higher orders in
the ERE.  The conformal invariance is not particularly illuminating in
single-channel scattering as unitarity constrains the one-dimensional $S$-matrix
trajectory to move in a one-dimensional space. However, in systems
where unitarity constrains the $S$-matrix to a surface with dimension greater
than one, the conformal invariance will be seen to provide a powerful constraint.

\subsubsection*{Nucleon-nucleon scattering: coordinates and fixed points}

\noindent The fixed points of the nucleon-nucleon $S$-matrix occur when the phase shifts both vanish,
$\delta_1=\delta_0=0$, or are both at unitarity,
$\delta_1=\delta_0=\frac{\pi}{2}$, or when $\delta_1=0$,
$\delta_0=\frac{\pi}{2}$ or $\delta_1=\frac{\pi}{2}$,
$\delta_0=0$. The $S$-matrices at these fixed points 
are the general solution of the equation $\hat {\bf S}^2=\hat {\bf 1}$
and are given by
\begin{eqnarray}
\hat {\bf S}_{\tiny{\circled{1}}}& =& +\hat {\bf 1}  \ \ \ , \ \ \ \hat {\bf S}_{\tiny{\circled{3}}}\ =\ +( \hat {\bf 1}+ \hat {\bm \sigma}\cdot \hat {\bm \sigma})/2 \ , \nn \\
\hat {\bf S}_{\tiny{\circled{2}}}& =& -\hat {\bf 1}    \ \ \ , \ \ \ \hat {\bf S}_{\tiny{\circled{4}}}\ = \ -( \hat {\bf 1}+ \hat {\bm \sigma}\cdot \hat {\bm \sigma})/2   \ .
   \label{eq:FPs}
\end{eqnarray}
These fixed points furnish a representation of the Klein four-group,
$\mathbb{Z}_2\otimes \mathbb{Z}_2$. As this is the discrete symmetry
group of the rhombus, the fixed points, which by construction provide the
boundaries of unitary interactions, suggest a geometrical
interpretation of the $S$-matrix as a trajectory within a rhombus
whose vertices are the fixed points of the RG and mark the most extreme values
that the $S$-matrix can achieve consistent with unitarity.

It is straightforward to make this geometrical construction precise.
In a $\mathbb{Z}_2$ basis consisting of ${\hat {\bf 1}}$ and $( \hat
{\bf 1}+ \hat {\bm \sigma}\cdot \hat {\bm \sigma})/2$, the $S$-matrix
can be decomposed as
\begin{eqnarray}
  \hat {\bf S}
  & = &
{1\over 2}\left(  e^{i 2 \delta_1} + e^{i 2 \delta_0} \right)
 \hat   {\bf 1}
\ +\
{1\over 2}\left( e^{i 2 \delta_1} - e^{i 2 \delta_0} \right)
\left(\hat {\bf 1}+ \hat  {\bm \sigma} \cdot   \hat  {\bm \sigma}\right)/2 \ .
   \label{eq:SdefB}
\end{eqnarray}
It is convenient to adopt the coordinates
\begin{eqnarray}
  \hat {\bf S} & = & u(p)\;\hat {\bf 1}\; +\; v(p)\;\left(\hat {\bf 1}+ \hat  {\bm \sigma} \cdot   \hat  {\bm \sigma}\right)/2 \ ,
   \label{eq:Sdefuv}
\end{eqnarray}
where $u(p)$ and $v(p)$ are complex functions decomposed in terms of real functions as
\begin{eqnarray}
u(p) \ =\ x(p)\;+\; i\;y(p) \ \ \ , \ \ \ v(p) \ =\ z(p)\;+\; i\;w(p) \ .
   \label{eq:xyztdef}
\end{eqnarray}
Unitarity provides two constraints on the four parameters; the first constraint
defines the unit three-sphere $S^3$:
\begin{eqnarray}
{\bar u}u\;+\; {\bar v}v\;=\;1 \;=\; x^2\;+\; y^2\;+\; z^2\;+\; w^2 \ ,
   \label{eq:S3def}
\end{eqnarray}
and the second constraint, 
\begin{eqnarray}
{\bar u}v+{\bar v} u \ =\ 0 \ =\ x z\; +\; y w \ ,
   \label{eq:S3const}
\end{eqnarray}
is a projective constraint whose intersection with $S^3$ provides the manifold on which the $S$-matrix
trajectory propagates. This choice of coordinates is by no means unique. A general parameterization of the $S$-matrix is
\begin{eqnarray}
  \hat {\bf S}   & = & \big\lbrack x(p)\;+\; i\;y(p) \big\rbrack    \hat   {\bf 1}
\ +\
\big\lbrack z(p)\;+\; i\;w(p) \big\rbrack
\left(\hat {\bf 1}\beta+ \hat  {\bm \sigma} \cdot   \hat  {\bm \sigma}\alpha\right) \ .
   \label{eq:Sdefgen}
\end{eqnarray}
With the coordinate choice
\begin{eqnarray}
  x & =&   {\textstyle\frac{1}{4\alpha}} \lbrack (3\alpha-\beta)\cos(\theta) + (\alpha+\beta)\cos(\phi) \rbrack \ \ , \ \ \nn \\
  y & =&   {\textstyle\frac{1}{4\alpha}} \lbrack (3\alpha-\beta)\sin(\theta) + (\alpha+\beta)\sin(\phi) \rbrack\ \ , \ \ \nn \\
  z & =&   {\textstyle\frac{1}{4\alpha}} \lbrack -\cos(\phi) + \cos(\theta) \rbrack\ \ , \ \ \nn \\
  w & =&   {\textstyle\frac{1}{4\alpha}} \lbrack -\sin(\phi) + \sin(\theta) \rbrack\ ,
   \label{eq:PSgen}
\end{eqnarray}
$\hat {\bf S}$ is independent of the parameters $\alpha$ and $\beta$. Here we have defined $\phi\equiv 2\delta_0$ and $\theta\equiv 2\delta_1$.
The unitarity constraints now take the form
\begin{eqnarray}
&& 1 \ =\ \left(x+\left(\alpha +\beta\right) z\right)^2\;+\; \left(y+\left(\alpha+\beta\right)w\right)^2 \ , \nn \\
&& \left(\alpha\;-\;\beta\right) \left( w^2\;+\;z^2\right) \ =\ x z\; +\; y w \ .
\label{eq:S3defgen}
\end{eqnarray}
Requiring that the coordinate system $(x,y,z,w)$ describe an isotropic space yields the
constraints $3\alpha-\beta=\alpha+\beta=1$, or $\alpha=\beta=1/2$ which recovers the 
choice made above in Eq.~(\ref{eq:Sdefuv}) and Eq.~(\ref{eq:xyztdef}), and leads to the parameterization of the $S$-matrix that will be
used throughout this paper\footnote{Another useful isotropic parameterization of the $S$-matrix is given by the Hopf-like coordinates, 
\begin{eqnarray}
  x & =& r \cos\xi \sin\eta \ \ , \ \ 
  y \ =\ r \sin\xi \sin\eta \ \ , \ \ \nn \\
  z & =& r \sin\xi \cos\eta \ \ , \ \ 
  w \ =\ -r \cos\xi \cos\eta \ . 
   \label{eq:Hopf}
\end{eqnarray}
with $\xi\in [0,2\pi)$ and $\eta\in [0,\pi]$.}:
\begin{eqnarray}
  x & =&   \oneht\;r \lbrack \cos(\phi) + \cos(\theta) \rbrack \ \ , \ \ 
  y \ =\   \oneht\;r \lbrack \sin(\phi) + \sin(\theta) \rbrack\ \ , \ \ \nn \\
  z & =&   \oneht\;r \lbrack -\cos(\phi) + \cos(\theta) \rbrack\ \ , \ \ 
  w \ =\   \oneht\;r \lbrack -\sin(\phi) + \sin(\theta) \rbrack\ ,
   \label{eq:PS}
\end{eqnarray}
with $\phi\in [0,2\pi]$ and $\theta\in [0,2\pi]$ and $r=1$. Note that the four coordinates span the range $[-1,1]$.
\begin{figure}[!ht]
\centering
\includegraphics[width = 0.75\textwidth]{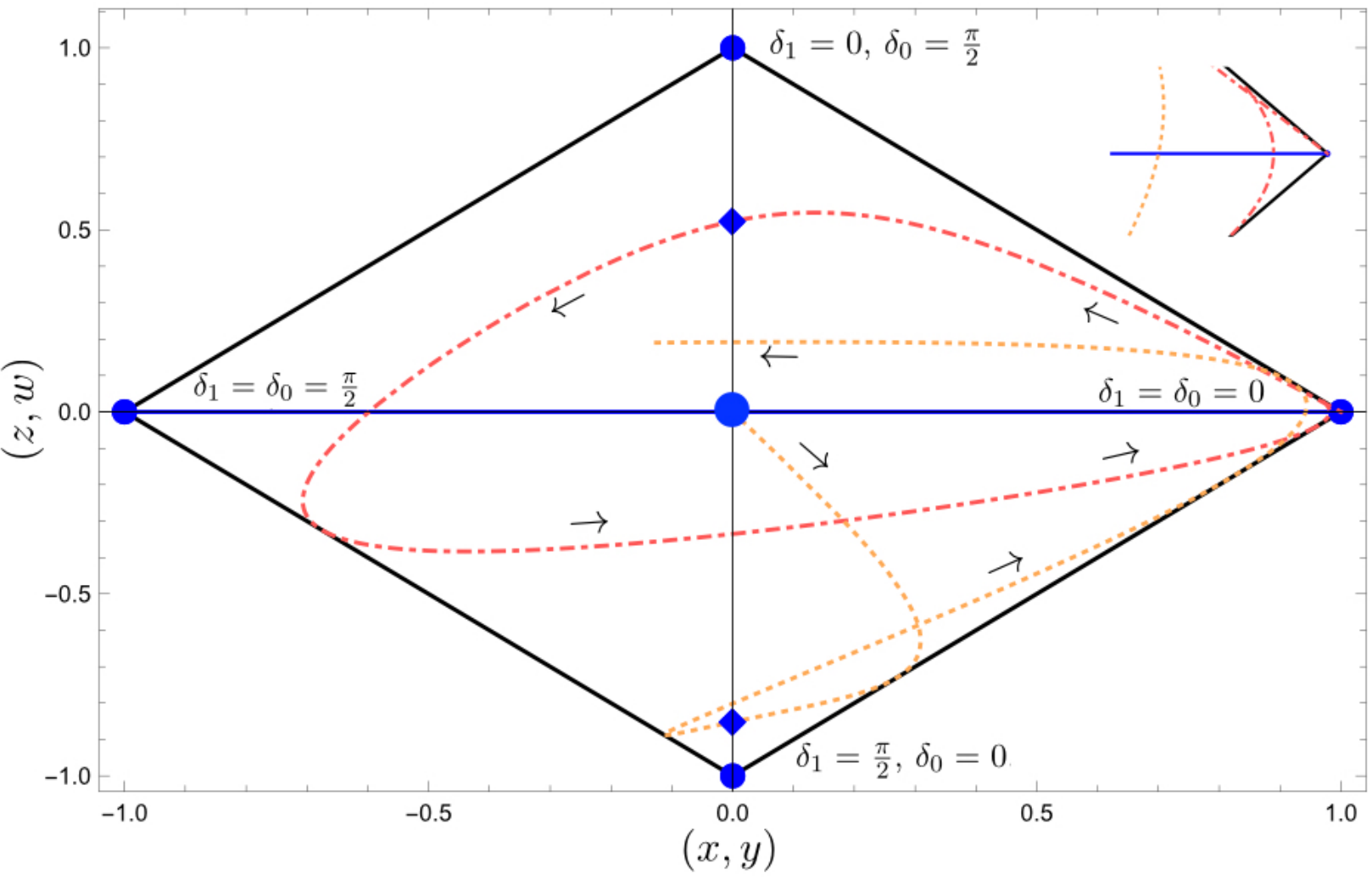}
\caption{Rhombus in the $u-v$ plane with fixed points (blue dots) at the vertices. The dot-dashed red curve corresponds to the
  real part of $\hat {\bf S}$ in the $x-z$ plane, and the dashed orange curve corresponds to the imaginary part of $\hat {\bf S}$ in the $y-w$ plane. These trajectories
  are obtained from the Nijmegen phase shift analysis (PWA93) in Ref.~\cite{NNOnline} and plotted from $p=0$ to $400~{\rm MeV}$ with the arrows indicating
  the direction of momentum flow. The blue diamonds correspond to the point at which the phase shifts differ by $\pi/2$ and the EP vanishes. The inset is a magnification of the right corner.}
  \label{fig:rhomb1}
\end{figure}

The fixed points of the RG in $(x,y,z,w)$ coordinates are:
\begin{eqnarray}
\hat {\bf S}_{\tiny{\circled{1}}}& =& (+1,\;0,\;0,\;0) \ \ ,\ \ \hat {\bf S}_{\tiny{\circled{3}}}\ =\ (\;0,\;0,+1,\;0) \ , \nn \\
\hat {\bf S}_{\tiny{\circled{2}}}& =& (-1,\;0,\;0,\;0)\ \ ,\ \ \hat {\bf S}_{\tiny{\circled{4}}}\ =\ (\;0,\;0,-1,\;0)\ .
   \label{eq:FPs2}
\end{eqnarray}

A geometrical description of scattering follows by mapping the
$\mathbb{Z}_2$ basis to the $u-v$ plane\footnote{In what follows, the
  $u-v$ plane will refer collectively to the $x-z$ and $y-w$ planes
  for the real and imaginary parts of the $S$-matrix, respectively.}
with $u$ representing the $\hat {\bf 1}$ axis and $v$ representing the
$( \hat {\bf 1}+ \hat {\bm \sigma}\cdot \hat {\bm \sigma})/2$
axis. The fixed points sit on the extrema of the axes in the $x-z$
plane. However, in the $y-w$ plane, the fixed points all sit at the
origin. In the plane defined by these new coordinates, the real and
imaginary parts of the $S$-matrix are confined to fall within the
boundaries set by $x+z=\pm 1$ and $x-z=\pm 1$, and $y+w=\pm 1$ and
$y-w=\pm 1$, respectively.  These boundaries define a rhombus in the
$u-v$ plane as shown in Fig.~(\ref{fig:rhomb1}).  The dihedral group
$D_2$, which is the symmetry group of the rhombus, corresponds to the
coordinate transformations $(u,v)\rightarrow (u,v)$ (identity),
$(u,v)\rightarrow (-u,v)$ (reflection about $v$), $(u,v)\rightarrow
(u,-v)$ (reflection about $u$), and $(u,v)\rightarrow (-u,-v)$
(rotation of axes by $\pi$). $D_2$ is isomorphic to the Klein
four-group, $\mathbb{Z}_2\otimes \mathbb{Z}_2$. From
Eq.~(\ref{eq:SdefB}), it is clear that when $\delta_0(p)=\delta_1(p)$,
the $S$-matrix is (trivially) symmetric with respect to the
$\mathbb{Z}_2$ subgroup $(u,v)\rightarrow (u,v)$ and $(u,v)\rightarrow
(u,-v)$ of the Klein group.  The nucleon-nucleon $S$-matrix taken from
scattering data in the form of the Nijmegen phase-shift
analysis~\cite{NNOnline} is plotted in Fig.~(\ref{fig:rhomb1}).

The projection of the $(x,y,z,w)$ coordinates onto two dimensions
allows a visualization of the real and imaginary parts of the
$S$-matrix. However as noted above, in the $y-w$ plane the fixed
points sit on top of each other whereas it is clear from
Eq.~(\ref{eq:FPs2}) that the four fixed points are distinct points in
the four dimensional space. Distinct locations of the fixed points is
achieved by projecting instead onto three of the four dimensions\footnote{
  The geometry of projections onto subspaces of the four-dimensional
  space will be considered in detail in the final section.}. However,
what is desired is a geometrical description of the $S$-matrix which
captures the four-dimensional nature of the space via the two independent
degrees of freedom given by the phase shifts $\phi$ and $\theta$.

\subsection{Entanglement power defined}

\noindent A direct consequence of the geometric picture of the
$S$-matrix as a trajectory confined to a rhombus is that when
$\delta_0(p)=\delta_1(p)$, the $S$-matrix trajectory is a line (geodesic) between fixed points of the RG and resides on a symmetry axis of the
rhombus that is protected by a $\mathbb{Z}_2$ subgroup of the Klein
four-group. In the EFT description this enhanced symmetry is
Wigner's $SU(4)$ spin-flavor symmetry where the two spin states of the
neutron and of the proton transform as the four-dimensional
fundamental
representation~\cite{Wigner:1936dx,Wigner:1937zz,Wigner:1939zz}.  This
symmetry also arises from the large-$N_c$ expansion in
QCD~\cite{Kaplan:1995yg,Kaplan:1996rk,CalleCordon:2008cz} and from the
near Schr\"odinger symmetry of the system implied by the large value
of the physical scattering lengths as compared to QCD length
scales~\cite{Mehen:1999qs}.  As this enhanced symmetry suggests a
suppression of spin-entangling interactions, there is motivation
to develop measures of entanglement that are suitable for
classifying interactions.

\begin{figure}[!ht]
\centering
\includegraphics[width = 0.65\textwidth]{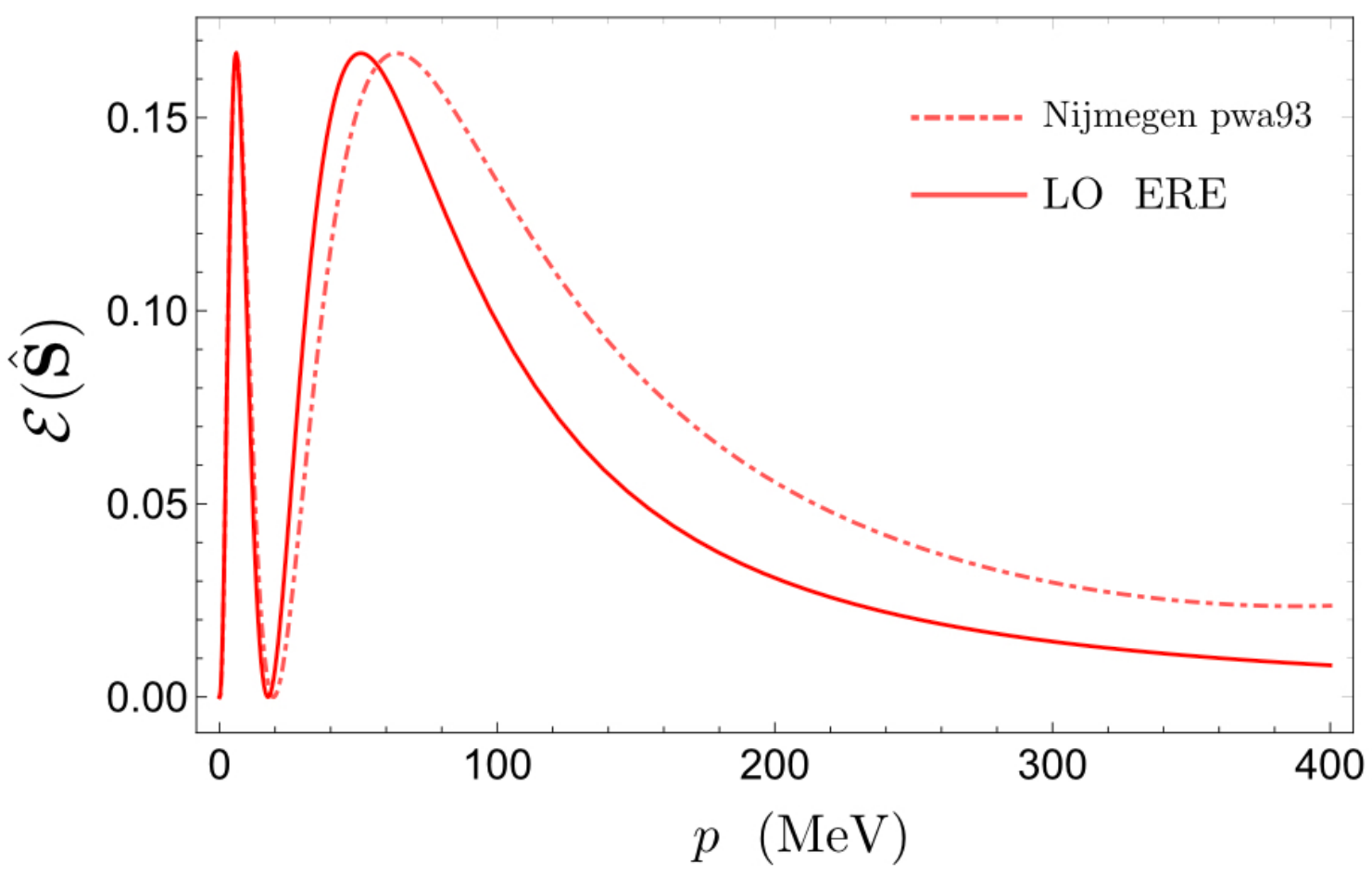}
\caption{The EP obtained from Eq.~(\ref{eq:epSnf2}) using the Nijmegen phase shift analysis in Ref.~\cite{NNOnline} (dot-dashed red curve),
from Eq.~(\ref{eq:epSscattapp}) using the scattering length only (red curve).}
    \label{fig:EPcurves}
\end{figure}

In a recent paper~\cite{Beane:2018oxh}, it was shown that the EP, 
a state-independent measure of quantum entanglement, can be defined
for the $S$-matrix:
\begin{equation}
{\mathcal E}({\hat {\bf S}})  = 1- \int {d\Omega_1\over 4\pi} \ {d\Omega_2\over 4\pi}\  {\rm Tr}_1\left[ \; \hat\rho_1^2 \;  \right]{\cal P}\left(\Omega_1,\Omega_2\right) \ ,
\label{eq:epgen}
\end{equation}
where $\hat\rho_1 = {\rm Tr}_2\left[\; \hat\rho_{12}\; \right]$ is the
reduced density matrix of the two-particle density matrix
$\hat\rho_{12} = |\psi_\text{out}\rangle\langle \psi_\text{out}| $ with $|\psi_\text{out} \rangle =
\hat {\bf S}|\psi_\text{in}\rangle$, and ${\cal P}$ is a probability distribution. The EP can be motivated as follows. Consider the non-entangled `in' state of the spin of two nucleons, $|\Omega_1 \rangle \otimes |\Omega_2 \rangle$, where $\Omega_{1,2}$ correspond to points on the Bloch sphere. After scattering, the outgoing state is generically no longer a product state and thus exhibits some amount of entanglement. The entanglement is quantified by the exponential of the second R\'enyi entropy, ${\rm Tr}_1\left[ \; \hat\rho_1^2 \;  \right]$. This non-unique entanglement measure is chosen as it is particularly simple to evaluate analytically. The final step is to sum the entanglement of the `out’ state for every possible `in’ state; this is the integration over the two Bloch spheres in Eq.~(\ref{eq:epgen}). This ensures that the EP is a state independent measure of entanglement and thus solely a property of the $S$-matrix.
For nucleon-nucleon scattering, the EP of $ \hat {\bf S}$ is
\begin{eqnarray}
{\mathcal E}({\hat {\bf S}})
& = &
{N}_{\cal P}\ \sin^2\left(2(\delta_1-\delta_0)\right) \ ,
  \label{eq:epSnf2}
\end{eqnarray}
where ${N}_{\cal P}$ is a numerical prefactor\footnote{With ${\cal
    P}=1$, ${N}_{\cal P}=1/6$. Choosing a distribution with more
  structure changes this prefactor.}. Expressed in terms of the coordinates
$(u,v)$ and $(x,y,z,w)$, one has
\begin{eqnarray}
{\mathcal E}({\hat {\bf S}})
& = & 4{N}_{\cal P}\ |u|^2  |v|^2 \ =\ 4{N}_{\cal P}\ \left( x^2+ y^2\right) \left( z^2+ w^2\right) \ ,
  \label{eq:epSnf2a}
\end{eqnarray}
and it is clear that the EP has geometrical significance
as a measure of length. (A concrete operator definition will be provided below.)
The EP vanishes
when $\delta_1(p)-\delta_0(p)= m {\pi\over 2}$ for any integer $m$.
This includes the $\mathbb{Z}_2$ symmetric line
$\delta_1(p)=\delta_0(p)$ as well as the one point at which the phase
shifts differ by $\pi/2$ (the blue diamond in
Fig.~(\ref{fig:rhomb1})). The EP is plotted for the
Nijmegen phase shift analysis~\cite{NNOnline} in
Fig.~(\ref{fig:EPcurves}). In addition, the EP vanishes
at the four fixed points of the renormalization group.
In general, the EP vanishes when
\begin{eqnarray}
  \hat {\bf S} \ =\ e^{i2\delta_0}\hat {\bf S}_{\tiny{\circled{1}}}\ \ \ {\rm or} \ \ \
  \hat {\bf S} \ =\ -e^{i2\delta_0}\hat {\bf S}_{\tiny{\circled{3}}}\ .
  \label{eq:epzeros}
\end{eqnarray}
Therefore, in the absence of EP, the $S$-matrix is characterized by
two curves that connect the fixed points $\hat {\bf S}_{\tiny{\circled{1}}}$ and $\hat {\bf S}_{\tiny{\circled{2}}}$
and $\hat {\bf S}_{\tiny{\circled{3}}}$ and $\hat {\bf S}_{\tiny{\circled{4}}}$ via the flow of the single phase shift $\delta_0(p)$
from $0$ to $\pi/2$.

Adapting momentum flow to the nucleon-nucleon system, consider the $S$-matrix under the action of a momentum dilatation $p \mapsto e^\epsilon p$. For small $\epsilon$,
\begin{eqnarray}
{\hat \beta({\hat {\bf S}})} \equiv p\frac{d}{dp} {\hat {\bf S}}(p) & = & \frac{1}{\epsilon}\left( {\hat {\bf S}}(e^\epsilon p)- {\hat {\bf S}}(p) \right) \ ,
\end{eqnarray}
and ${\hat \beta({\hat {\bf S}})}=0$ at the four fixed points where ${\hat {\bf S}}^2=\hat {\bf 1}$.
In the $\mathbb{Z}_2$ basis one then has
\begin{eqnarray}
  {\hat \beta({\hat {\bf S}})} &=& {\beta_u(p)}\; \hat {\bf 1} \ +\ {\beta_v(p)} \left(\hat {\bf 1}+ \hat  {\bm \sigma} \cdot   \hat  {\bm \sigma}\right)/2 \ ,
\end{eqnarray}
with
\begin{eqnarray}
{\beta_u(p)} \equiv p\frac{d}{dp} u(p) \ \ \ , \ \ \ {\beta_v(p)} \equiv p\frac{d}{dp} v(p) \ .
\end{eqnarray}
Eq.~(\ref{eq:epSnf2a}) makes clear that the EP is related to the distance of the $u$ and $v$ coordinates from the origin.
However, given that the EP has support only away from the fixed points, one might expect that when the momentum
dependence of the $S$-matrix is specified, the EP will be directly related to ${\beta_{u,v}(p)}$.
Indeed, generalizing the expression for $u v$ given in
Eq.~(\ref{eq:epSscattappcurvemod2}) to a linear combination of $\beta_u$
and $\beta_v$ with complex coefficients, it is straightforward to find
an expression for the EP in terms of the beta functions alone
\begin{eqnarray}
  {\mathcal E}({\hat {\bf S}}) & = & \frac{N_{\cal P}}{4} \bigg\lvert\; \frac{\beta_{\bar v}(p) \beta_{u}(p) - \beta_{\bar u}(p) \beta_{v}(p)}{\beta^2_{\bar u}(p) - \beta^2_{\bar v}(p)} \; \bigg\rvert^2 \;.
  \label{eq:epgeneral} 
\end{eqnarray}
The EP expressed in this form exhibits its connection to the RG; scattering at the RG fixed points does not entangle spins.

\subsection{Effective range theory at leading order}

\begin{figure}[!ht]
\centering
\includegraphics[width = 0.75\textwidth]{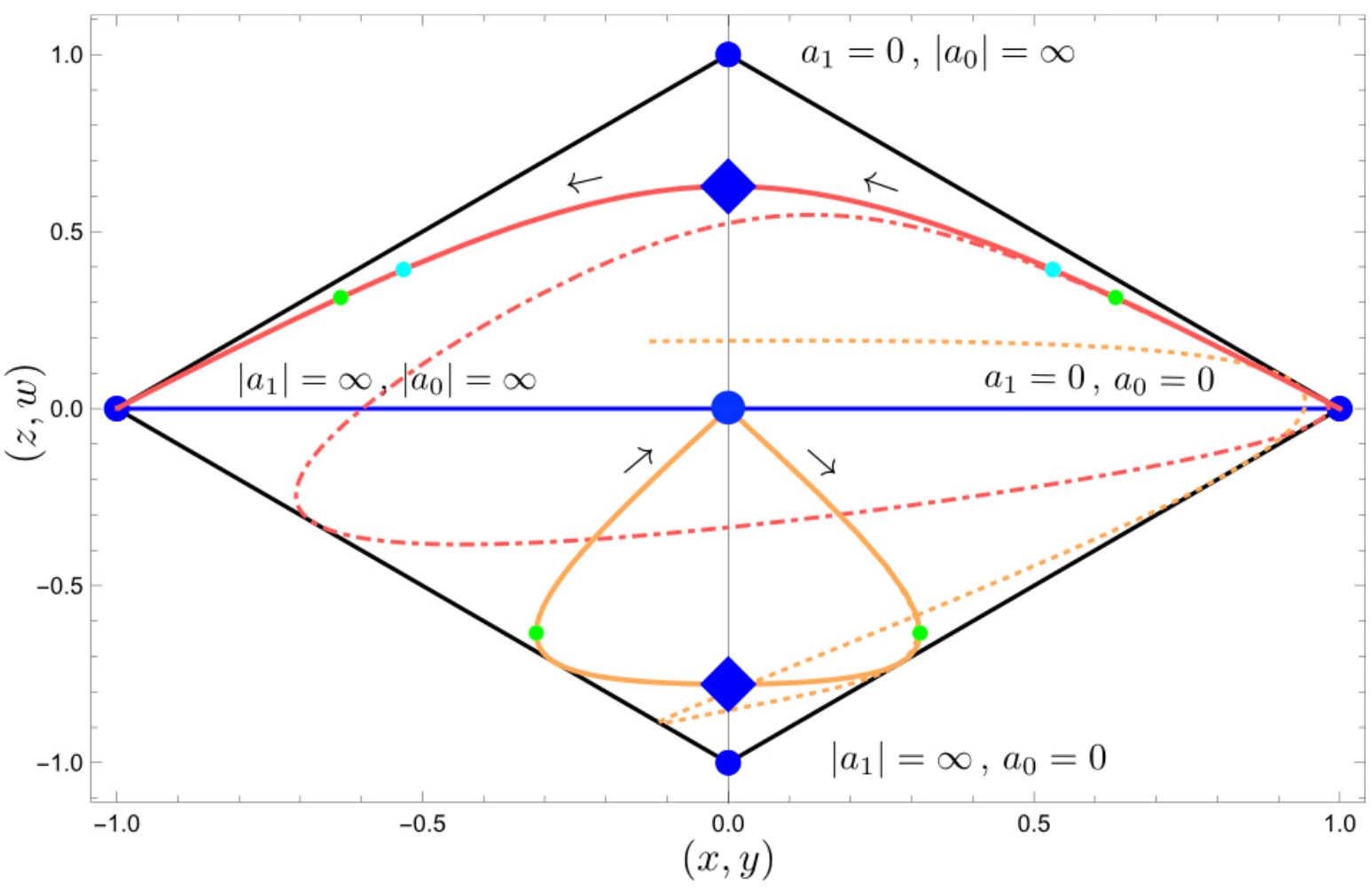}
\caption{Rhombus in the $u-v$ plane with fixed points at the vertices and the origin. The solid red curve corresponds to the
  real part of $\hat {\bf S}$, and the solid orange curve corresponds to the imaginary part of $\hat {\bf S}$ at LO in the ERE,
  with the physical values of the nucleon-nucleon scattering lengths. The Nijmegen phase shift analysis~\cite{NNOnline} is also shown for comparison.
  The blue diamonds correspond to the fixed point of the UV/IR conformal transformation. The green dots are the maxima of the EP and the cyan dots are the maxima of a measure
  of curvature, as described in the text.}
  \label{fig:rhomb2}
\end{figure}

\noindent At LO in the ERE, the $S$-matrix is completely determined by the scattering lengths and
is expressed in the chosen coordinate basis as
\begin{eqnarray}
  u(p) & =& \frac{1}{2}\left( \frac{1- i a_1 p}{1+ i a_1 p}+\frac{1- i a_0 p}{1+ i a_0 p} \right) \ \ \   , \  \ \
  v(p) \ =\ \frac{1}{2}\left( \frac{1- i a_1 p}{1+ i a_1 p}-\frac{1- i a_0 p}{1+ i a_0 p} \right) \ .
   \label{eq:Sdef3}
\end{eqnarray}
In terms of phase shifts,
\begin{eqnarray}
\phi & =& -2\tan^{-1}\!\left( a_0 p\right) \ \ \ , \ \ \ \theta\ =\ -2\tan^{-1}\!\left( a_1 p\right) \ .
 \label{eq:LOinaffineSOL}
\end{eqnarray}
The fixed points of the RG are now at $a_1= a_0=0$,
$|a_1|=| a_0|=\infty$, and at $a_1=0\,,\, |a_0|=\infty$, and
$|a_1|=\infty\,,\, a_0=0$.  In nucleon-nucleon scattering, the
scattering lengths are determined experimentally to be
$a_0=-23.714\pm0.013~{\rm fm}$ and $a_1=5.425\pm 0.002~{\rm fm}$. As
seen in Fig.~(\ref{fig:rhomb2}), Eq.~(\ref{eq:Sdef3}) provides a good
approximation to the physical nucleon-nucleon $S$-matrix for $p\leq
10~{\rm MeV}$, and, of course, this corresponds to leading-order in
the EFT description.
Now assume that all inelastic thresholds are pushed
to infinity so that $p$ is defined on the interval $[0,\infty)$, and further assume
that all range and shape parameters vanish. These assumptions
extend the LO ERE to arbitrarily high momenta. In the EFT, this corresponds to a theory with only momentum-independent contact interactions between nucleons; i.e. all derivative interactions are assumed to vanish. By construction this model maps onto the correct physics in the IR and reveals interesting properties of the $S$-matrix when extended to the UV.

The real and imaginary parts of the $S$-matrix, described by $x$ and
$z$, and by $y$ and $w$, respectively, are in correspondence with
distinct trajectories in the $u-v$ plane that propagate between fixed
points. A striking feature of Fig.~(\ref{fig:rhomb2}) is the
isometries of the $S$-matrix trajectories at LO in the ERE when the
momenta ranges over the entire positive half line; for the physical
scattering lengths, both the real and imaginary trajectories of $S$
are $(u,v)\rightarrow (-u,v)$ symmetric. In order to analyze the
isometries systematically, it is important to distinguish between the
various cases which are determined by the relative signs of the
scattering lengths. We will refer to the scenario where $a_1 a_0 <0$
as ``singly-bound'' (one channel bound and the other unbound as in the
physical case) and the scenario where $a_1 a_0 >0$ as
``doubly-(un)bound'' (both channels bound or both channels
unbound). The symmetry of the $S$-matrix and the EP are strikingly
different in these two scenarios.

The isometries of the $S$-matrix trajectories are in correspondence with the UV/IR  transformation on the momenta,
\begin{eqnarray}
p\mapsto \frac{1}{|a_1 a_0| p} \ .
   \label{eq:moebius}
\end{eqnarray}
This invariance, generalized from single-channel scattering, is an intrinsic property of the LO ERE $S$-matrix at
finite, non-zero values of the scattering lengths. That this symmetry is not
visible in the LO EFT action is no surprise;
the momentum range of the $S$-matrix has been extended to encompass the entire 
positive half line, and the symmetry interchanges the UV and the IR. Therefore, the 
$S$-matrix at LO in the ERE is a UV complete description of scattering, in sharp contrast with the LO EFT description.
The two fixed points of the
conformal transformation are at $p=\pm \sqrt{|a_0a_1|^{-1}}$.
Therefore, for a given trajectory, only one of the fixed
points of the transformation appears in the physical (scattering)
region ($p\geq 0$) and lies on the axis of reflection of the corresponding isometry
(the diamonds in Fig.~(\ref{fig:rhomb2})).

Acting with Eq.~(\ref{eq:moebius}) on Eq.~(\ref{eq:Sdef3}), it is straightforward to find the UV/IR transformation on the coordinates and angles:
\begin{eqnarray}
&& {\underline{a_1 a_0 <0}}:\  (x,z)\rightarrow (-x,z) \ \ , \ \ (y,w)\rightarrow (-y,w) \Longrightarrow \ (u,v)\rightarrow (-u,v) \ , \nn \\
&&\qquad\qquad\qquad\qquad (\phi,\theta)\rightarrow(\theta\mp\pi,\phi\pm\pi) \ , \nn \\
  &&{\underline{a_1 a_0 >0}}:\  (x,z)\rightarrow (-x,z) \ \ , \ \ (y,w)\rightarrow (y,-w) \Longrightarrow \ (u,v)\rightarrow (-\bar u,\bar v) \ , \nn \\
&&\qquad\qquad\qquad\qquad (\phi,\theta)\rightarrow(-\theta\pm\pi,-\phi\pm\pi) \ .
  \label{eq:moebiusiso2}
\end{eqnarray}
Thus, the UV/IR transformation is a conformal transformation which leaves linear combinations of phase shifts invariant:
\begin{eqnarray}
&& {\underline{a_1 a_0 <0}}:\  \phi+\theta \rightarrow \phi+\theta\nn \\
  &&{\underline{a_1 a_0 >0}}:\  \phi-\theta \rightarrow \phi-\theta \ .
  \label{eq:confmoebiusiso2}
\end{eqnarray}
This conformal symmetry will prove critical in what follows.

\begin{figure}[!ht]
\centering
\includegraphics[width = 0.75\textwidth]{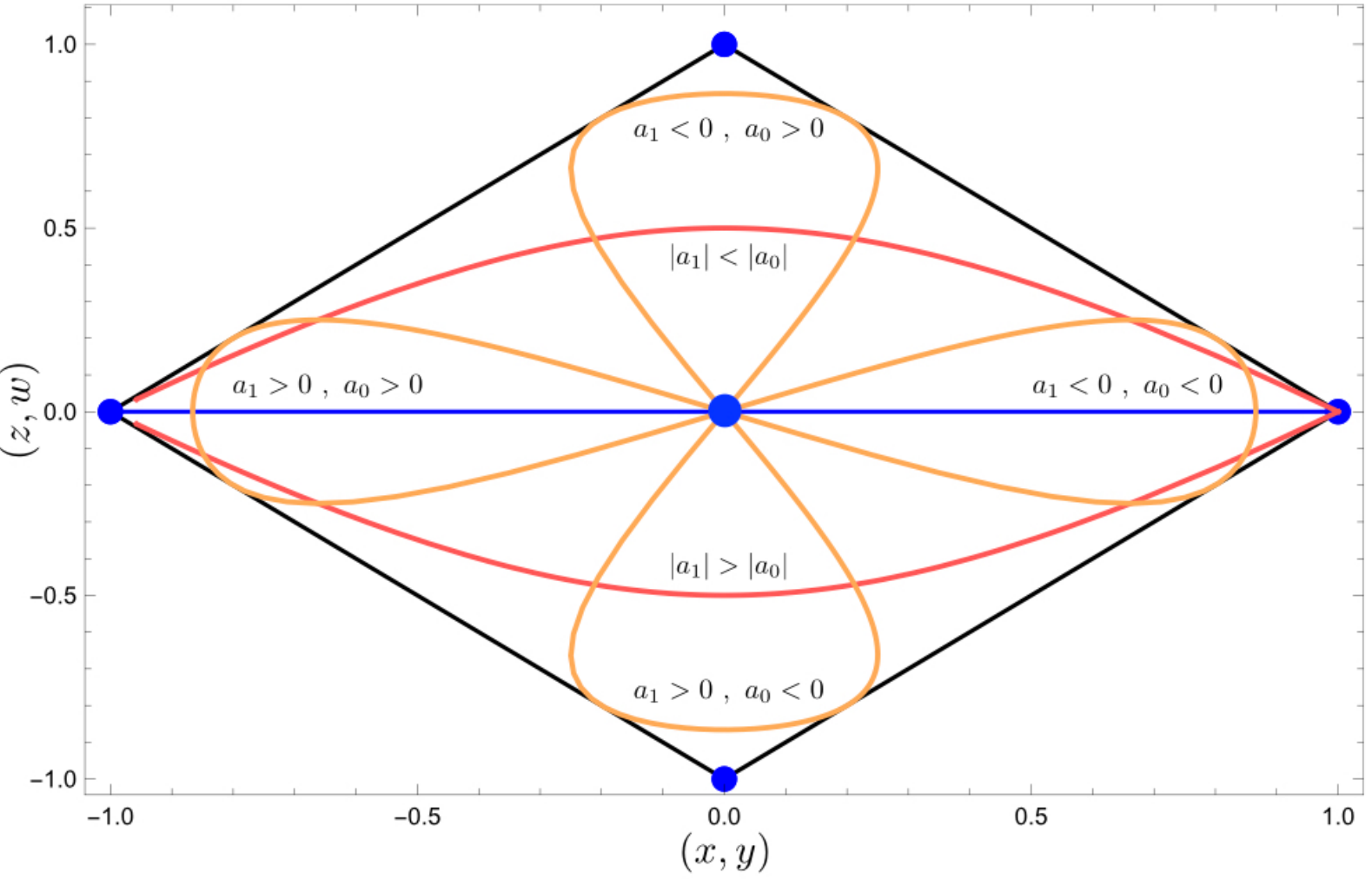}
\caption{Rhombus in the $u-v$ plane with fixed points at the vertices and the origin. The solid red curves correspond to the
  real part of $\hat {\bf S}$, and the solid orange curves correspond to the imaginary part of $\hat {\bf S}$ at LO in the ERE. $S$-matrix trajectories with scattering lengths
  whose magnitudes are fixed to the physical values are displayed with all relative sign combinations.}
  \label{fig:rhomb3}
\end{figure}

For an illustration of the isometries in the $u-v$ plane with the scattering length magnitudes fixed to the physical case but with signs flipped, see Fig.~(\ref{fig:rhomb3}).

At LO in the ERE, the beta functions for the coordinates are easily found to be
\begin{eqnarray}
p\frac{d}{dp} v(p) & = & u(p) v(p) \ \ \ , \ \ \ p\frac{d}{dp} u(p) \ = \ \oneht\left( u(p)^2 + v(p)^2 - 1\right) \ .
\label{eq:vdiffeq} 
\end{eqnarray}
And consequently the momentum-flow equation for the $S$-matrix is
\begin{eqnarray}
 p\frac{d}{dp} {\hat {\bf S}} & = & \oneht \left( {\hat {\bf S}}^2 \;-\; \hat {\bf 1} \right) \ .
\end{eqnarray}
As in the single-channel case, this takes the form of the Ward identity for dilatations and therefore vanishes at the fixed
points of the RG, as required. The content of this equation for phase shift degrees of freedom is a pair
of decoupled, first-order, differential equations of the simple form $p\,\dot\phi=\sin\phi$ and
$p\,\dot\theta=\sin\theta$ which immediately integrate to the solution of Eq.~(\ref{eq:LOinaffineSOL}).

The EP of $ \hat {\bf S}$ in terms of scattering lengths is
\begin{eqnarray}
{\mathcal E}({\hat {\bf S}})
& = &
4 N_{\cal P}\frac{p^2\left( a_0-a_1\right)^2 \left(1+a_1 a_0 p^2\right)^2}{\left(1+a_0^2 p^2\right)^2\left(1+a_1^2 p^2\right)^2 } \ .
\label{eq:epSscattapp} 
\end{eqnarray}
Note that there is a zero of the EP only in the singly-bound case.
The EP is conformally invariant and has three (non-trivial) extrema as
is evident in Fig.~(\ref{fig:EPcurves}) in the physical case; there is a minimum at the
conformal fixed point and two local maxima (shown as green dots on the
$S$-matrix trajectory in Fig.~(\ref{fig:rhomb2})).  The EP therefore
behaves like the curvature of the $S$-matrix trajectory in the
interior of the rhombus.  A conformally invariant measure of curvature
can be constructed from the square of the $z(p)$ coordinate beta
function.  This shares the bulk features of the EP in the singly-bound
case, as shown in Fig.~(\ref{fig:EPcurves}) (cyan dots), where
$N_{\cal P}$ has been adjusted so that the maxima overlap.  The EP
must depend also on the imaginary trajectory as this contains
information about the signs of the scattering lengths. Indeed, one
readily finds that at LO in the ERE,
\begin{eqnarray}
{\mathcal E}({\hat {\bf S}})
& = &
4 N_{\cal P}  \big\lvert\; \beta_v(p)\; \big\rvert^2 \ .
\label{eq:epSscattappcurve} 
\end{eqnarray}
Writing
\begin{eqnarray}
{\mathcal E}({\hat {\bf S}})
& = &
{\mathcal E}( {\text Re}{\hat {\bf S}}) \ +\ {\mathcal E}({\text Im}{\hat{\bf S}}) \ ,
\label{eq:epSscattappcurve2} 
\end{eqnarray}
with
\begin{eqnarray}
  {\mathcal E}( {\text Re}{\hat {\bf S}}) \ =\  N_{\cal P} \left( p\frac{d}{dp} z(p)\right)^2 \ \ , \ \
  {\mathcal E}({\text Im}{\hat{\bf S}})  \ =\  N_{\cal P} \left( p\frac{d}{dp} w(p)\right)^2 \;,
\label{eq:epSscattappcurve3} 
\end{eqnarray}
one sees that there is a distinct contribution to the EP from each of the trajectories, both of which are related
to the curvature of the $v(p)$ coordinate.

\subsection{Conformal range model}

\begin{figure}[!ht]
\centering
\includegraphics[width = 0.75\textwidth]{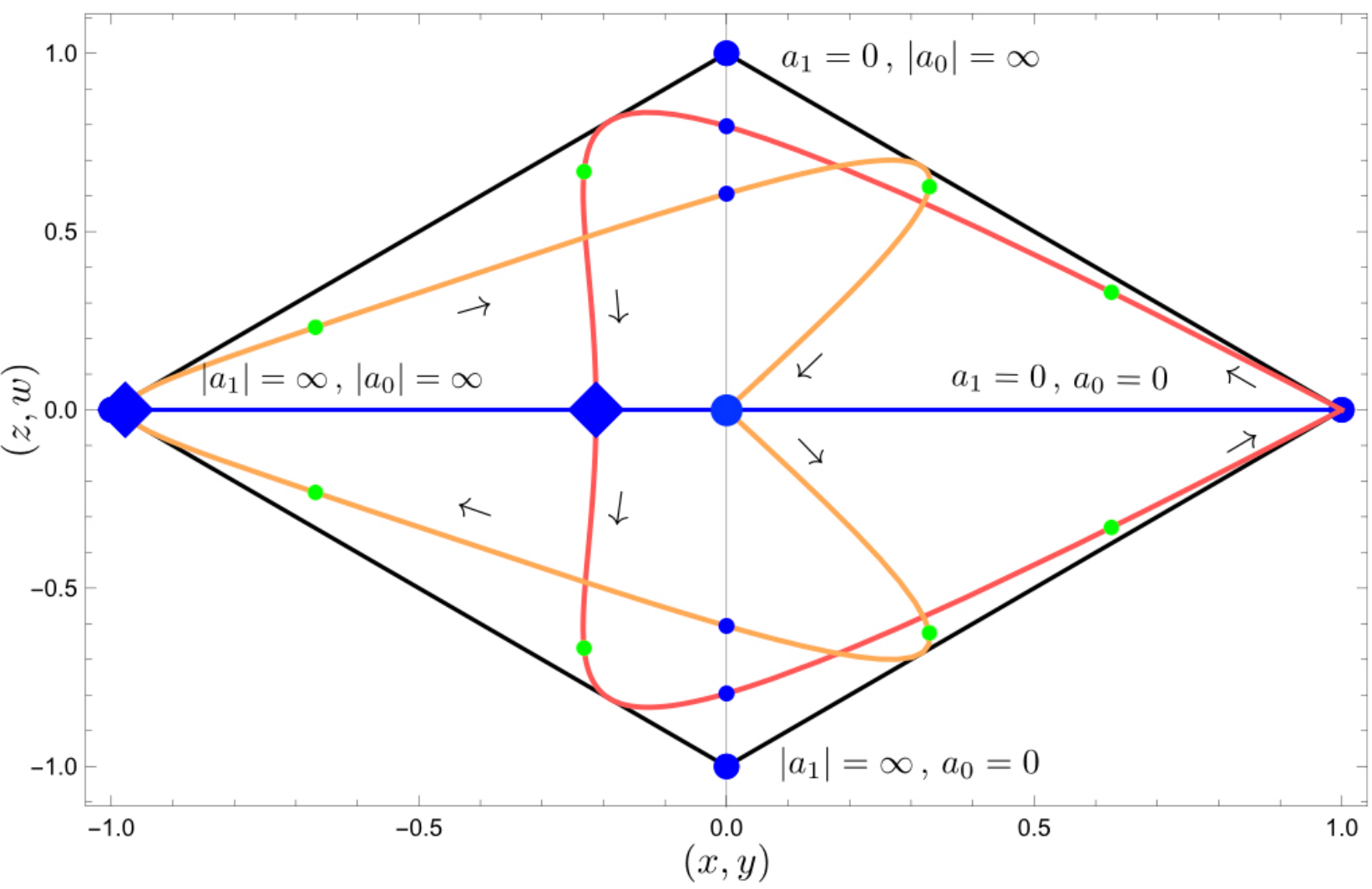}
\caption{Rhombus in the $u-v$ plane with fixed points at the vertices and the origin. The solid red curve corresponds to the
  real part of $\hat {\bf S}$, and the solid orange curve corresponds to the imaginary part of $\hat {\bf S}$ at NLO in the ERE,
  with the physical values of the nucleon-nucleon scattering lengths and $\lambda= \frac{1}{4}$. 
  The blue diamonds correspond to the fixed point of the UV/IR conformal transformation. The green (blue) dots are the maxima (minima) of the EP.}
    \label{fig:rhombusmod2}
\end{figure}

\noindent Generally, the conformal invariance which interchanges the UV and the IR will be broken beyond LO in the ERE.
Higher orders in the ERE include more short-distance structure, and maintaining a UV/IR symmetry will require
contrivance\footnote{Note that the $S$-matrix is, by construction, unitary to all orders in the ERE. While it is sometimes convenient in low-energy nucleon-nucleon scattering to treat effective range corrections in perturbation theory~\cite{Kaplan:1998tg,Kaplan:1998we} ---which leads to an $S$-matrix that satisfies perturbative unitarity--- this is not done here.}. Nevertheless, it is straightforward to find $S$-matrix models that include higher orders in the ERE and 
which possess UV/IR symmetry. These models are interesting as the stronger momentum dependence implies $S$-matrix trajectories
with more complex curvature and EP. Consider
\begin{eqnarray}
\hspace{-0.2in}  u(p) & =& \frac{1}{2}\left( \frac{1- i a_1(p) p}{1+ i a_1(p) p}+\frac{1- i a_0(p)p}{1+ i a_0(p) p} \right) \ , \ 
  v(p) \ =\ \frac{1}{2}\left( \frac{1- i a_1(p) p}{1+ i a_1(p) p}-\frac{1- i a_0(p) p}{1+ i a_0(p) p} \right) ,
   \label{eq:Sdefuvap}
\end{eqnarray}
where the momentum-dependent scattering length is defined as
\begin{eqnarray}
a_r(p) \ =\ \frac{a_r}{1-\lambda a_r^2 p^2} \ ,
   \label{eq:mdepscatt}
\end{eqnarray}
and $\lambda>0$ is a real number.  This model reduces to the LO ERE
when $\lambda=0$, and with $\lambda\neq 0$
includes effective range corrections\footnote{Note that in the
  singly-bound case, only one effective range will be positive in this
  model, in violation of the Wigner bound~\cite{Wigner:1955zz,Phillips:1996ae,Hammer:2010fw}.}.
As in the LO ERE, the real and imaginary parts of the $S$-matrix
may be viewed as trajectories in the $u-v$ plane. These $S$-matrix
trajectories in the rhombus with physical values of the
scattering lengths and with the choice $\lambda= \frac{1}{4}$ are shown in
Fig.~(\ref{fig:rhombusmod2}).  The trajectories in this model are
significantly more complex than the LO case and yet have clear isometries.

Here the isometries of the $S$-matrix trajectories are in correspondence with the UV/IR transformation on the momenta,
\begin{eqnarray}
p\mapsto \frac{1}{\lambda |a_1 a_0| p} \ .
   \label{eq:moebiusallorders}
\end{eqnarray}
The action of this transformation on the coordinates and angles is 
\begin{eqnarray}
&& {\underline{a_1 a_0 <0}}:\  (x,z)\rightarrow (x,-z) \ \ , \ \ (y,w)\rightarrow (y,-w) \Longrightarrow \ (u,v)\rightarrow (u,-v) \ , \nn \\
&&\qquad\qquad\qquad\qquad (\phi,\theta)\rightarrow(\theta,\phi) \ , \nn \\
  &&{\underline{a_1 a_0 >0}}:\  (x,z)\rightarrow (x,-z) \ \ , \ \ (y,w)\rightarrow (-y,w) \Longrightarrow \ (u,v)\rightarrow (\bar u,-\bar v) \ , \nn \\
&&\qquad\qquad\qquad\qquad (\phi,\theta)\rightarrow(-\theta,-\phi) \ .
  \label{eq:moebiusallordersiso2}
\end{eqnarray}
And the corresponding conformal invariance is again as in Eq.~(\ref{eq:confmoebiusiso2}).

A straightforward application of the chain rule leads to the beta functions of the coordinates
\begin{eqnarray}
  p\frac{d}{dp} u(p) & = & \oneht \Big\lbrack \left( \frac{a_1(p)}{a_1}+ \frac{a_0(p)}{a_0}-1 \right)\left( u(p)^2 + v(p)^2 - 1\right) +
  2\left( \frac{a_1(p)}{a_1}- \frac{a_0(p)}{a_0} \right) u v \Big\rbrack  \ ; \\
    p\frac{d}{dp} v(p) & = & \oneht \Big\lbrack \left( \frac{a_1(p)}{a_1}- \frac{a_0(p)}{a_0} \right)\left( u(p)^2 + v(p)^2 - 1\right) +
2\left( \frac{a_1(p)}{a_1}+ \frac{a_0(p)}{a_0}-1 \right) u v \Big\rbrack  \ .
\label{eq:vdiffeqmod2} 
\end{eqnarray}
And consequently $S$-matrix momentum flow is given by
\begin{eqnarray}
  p\frac{d}{dp} {\hat {\bf S}} & = & \oneht \left( {\hat {\bf S}}^2 \;-\; \hat {\bf 1} \right)
  \Big\lbrack \left( \frac{a_1(p)}{a_1}+ \frac{a_0(p)}{a_0}-1 \right)  \hat {\bf 1}   +
  \left( \frac{a_1(p)}{a_1}- \frac{a_0(p)}{a_0} \right)
\left(\hat {\bf 1}+ \hat  {\bm \sigma} \cdot   \hat  {\bm \sigma}\right)/2 
  \Big\rbrack  \ .
\end{eqnarray}
This momentum-flow equation reduces to LO in the ERE when $\lambda=0$.
\begin{figure}[!ht]
\centering
\includegraphics[width = 0.65\textwidth]{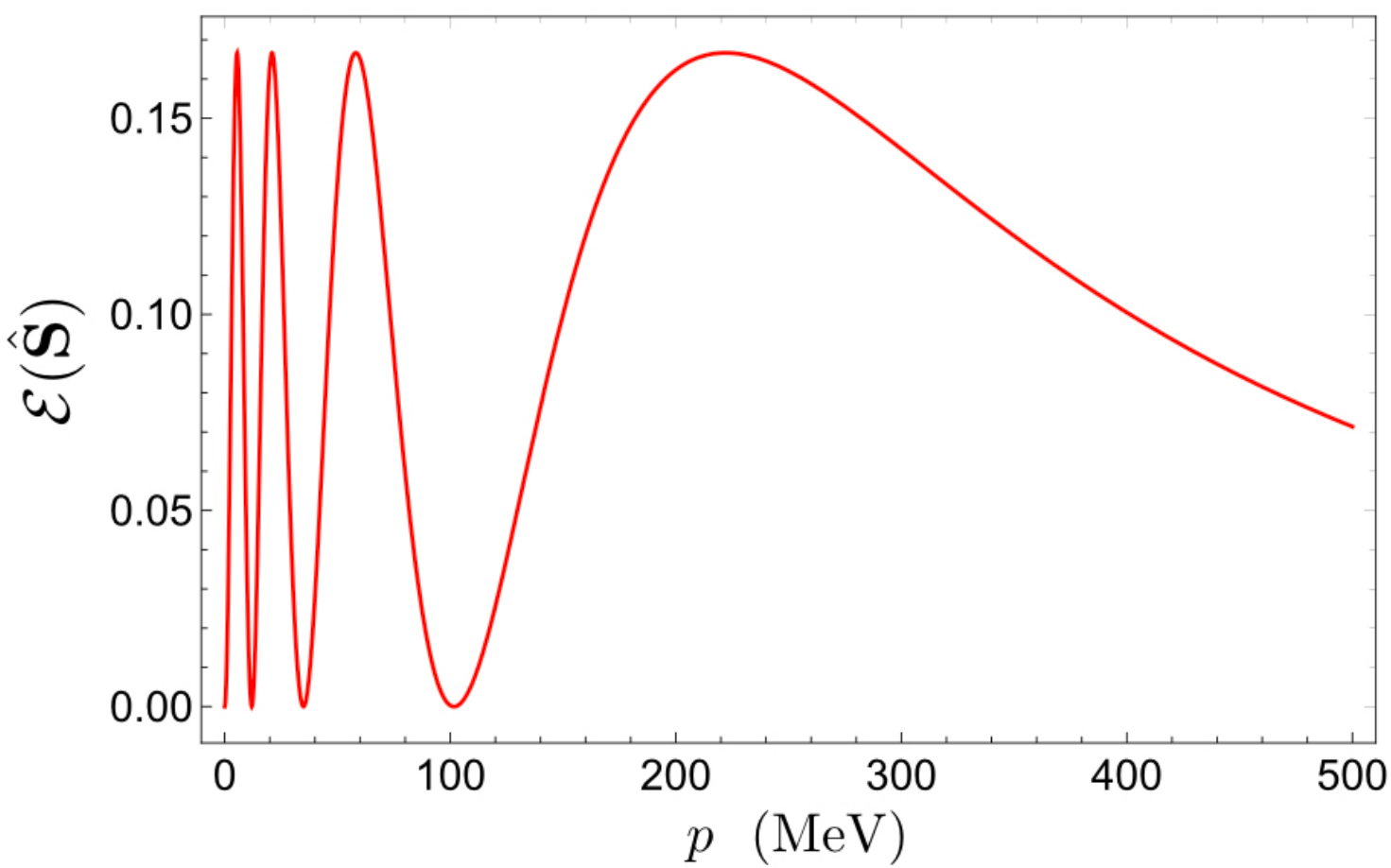}
\caption{The EP obtained from Eq.~(\ref{eq:epSnf2mod2}) with the physical values of the scattering lengths and $\lambda= \frac{1}{4}$.}
    \label{fig:EPcurves2}
\end{figure}

The EP is given by
\begin{eqnarray}
\hspace{-0.3in} {\mathcal E}({\hat {\bf S}})
& = & 4 N_{\cal P}
\frac{p^2\left( a_0-a_1\right)^2 \left(1+a_1 a_0 \lambda p^2\right)^2\big\lbrack 1+p^2\left(a_0 a_1-\left(a_1^2+a_0^2\right)\lambda+ a_1^2 a_0^2 \lambda^2 p^2\right)\big\rbrack^2}
     {\big\lbrack 1+a_0^2 p^2\left(1-\lambda\left(2-a_0^2 \lambda p^2\right)\right) \big\rbrack^2 \big\lbrack 1+a_1^2 p^2\left(1-\lambda\left(2-a_1^2 \lambda p^2\right)\right) \big\rbrack^2 } \ ,
\label{eq:epSnf2mod2}
\end{eqnarray}
and can be written in terms of the $S$-matrix beta functions as
\begin{eqnarray}
{\mathcal E}({\hat {\bf S}})
& = &
4 N_{\cal P} \big\lvert\; c_- (p)\; \beta_u(p)  \ +\ c_+ (p)\;\beta_v(p)\; \big\rvert^2 \ ,
\label{eq:epSscattappcurvemod2} 
\end{eqnarray}
where
\begin{eqnarray}
  c_\pm (p) \equiv \pm
  \frac{\big\lbrack\left(2a_0 a_1(p)-a_1a_0\right)\pm \left(2a_1 a_0(p)-a_1a_0\right)\big\rbrack}{2\left(2a_1(p)-a_1\right)\left(2a_0(p)-a_0\right)} \ .
\end{eqnarray}
The EP is plotted in Fig.~(\ref{fig:EPcurves2}) with the physical values of the scattering lengths and $\lambda= \frac{1}{4}$. Again the
EP is conformally invariant. With physical values of the scattering lengths, there are now three zeros of the EP corresponding to the crossing of the fixed point of the conformal transformation and the crossing of the $u=0$
line of zero entanglement.

\section{Geometry of the $S$-matrix}
\label{sec:geosmatse}

\subsection{Hilbert-Schmidt Distance}

\noindent The analysis of Section~\ref{sec:smatse} relies on 
choosing a specific isotropic coordinate system to study the geometry
of the $S$-matrix. As the $S$-matrix is an operator in the product
Hilbert space of nucleon spins, it is interesting to consider distance
measures in a basis-independent manner.  For this purpose it is
convenient to make use of the Hilbert-Schmidt (HS) distance.  The HS
distance measure is a natural extension of the Frobenius inner
product, $\langle \hat{\bf A},\hat{\bf B} \rangle = {\rm Tr}\lbrack
\hat{\bf{ A}^\dagger} \hat{\bf{ B}} \rbrack$. It can be defined as~\cite{bengtsson_zyczkowski_2006}
\begin{equation}
    D(\hat{\bf{A}},\hat{\bf{B}})^2  \equiv
    d_n{\rm Tr}\left[\; (\hat{\bf{A}}-\hat{\bf{B}})(\hat{\bf{A}}-\hat{\bf{B}})^\dagger\; \right]
    \label{eq:hsdist}
\end{equation}
with $d_n$ an arbitrary normalization constant that will be set to $\frac{1}{2}$.
The HS distance is independent of basis, positive semi-definite and
zero if and only if $\hat{\bf{A}} = \hat{\bf{B}}$. If the $S$-matrix is
parameterized by phase shifts, say $\phi$ and $\theta$, then
the HS distance induces a metric on the space of $S$-matrices. This
allows for the direct study of the geometry of the S-matrix. The HS
distance between two $S$-matrices with distinct phase shifts, $\hat{\bf
  S}(\phi, \theta)$ and $\hat{\bf S'}(\phi',\theta')$, is
\begin{equation}
    D(\hat{\bf S}, \hat{\bf S'})^2 = 
     \oneht{\rm Tr}\left[\; (\hat{\bf S} - \hat{\bf S'})(\hat{\bf S} - \hat{\bf S'})^\dagger\; \right] = 2 \Big(\sin^2\left(\oneht(\phi-\phi')\right) + 3\sin^2\left(\oneht(\theta-\theta')\right) \Big) \ .
    \label{eq:globalmet}
\end{equation}
The metric is obtained by looking at the infinitesimal differences, $d\phi = \phi' - \phi$ and $d\theta = \theta' - \theta$ and is found to be,
\begin{equation}
    ds^2 = \oneht\left(3\,d \theta^2 + d\phi^2\right) \ .
    \label{eq:localmet}
\end{equation}
The unitary $S$-matrix is determined by the two degrees of freedom, $\phi$ and $\theta$,
and therefore, locally, the $S$-matrix lives on the space defined by this two-dimensional Euclidean metric
that can be rescaled to remove the anisotropic spin weighting factor of the spin-triplet phase shift $\theta$.
A more geometrical approach to obtaining this metric, using an embedding, will be pursued in the next section.

The HS distance allows a definition of the analog of the ``Hamiltonian operator'' which governs
momentum flow of the $S$-matrix. As a result of unitarity, two $S$-matrices, $\hat{\bf{S}}(p)$ and $\hat{\bf{S'}}(p+dp)$, that are infinitesimally
close to each other in the two-dimensional Euclidean space are related by,
\begin{equation}
     \hat{\bf{S'}} =  \hat{\bf{S}} (\hat {\bf 1} + i \hspace{.06cm} dp \hspace{.06cm} \hat{\bf{H}}) \ ,
     \label{eq:hermittrans}
\end{equation}
where $\hat{\bf{H}}$ is a Hermitian matrix. The differential line element between $\hat{\bf{S}}$ and $\hat{\bf{S'}}$ can be expressed in terms of the Frobenius norm as,
\begin{equation}
    ds^2  = \oneht dp^2   \langle \hat{\bf H},\hat{\bf H} \rangle \ .
    \label{eq:althsmetric}
\end{equation}
Therefore, the arc-length along a curve is
\begin{equation}
    \frac{1}{2} \int dp \sqrt{\hspace{.1cm} \langle \hat{\bf H}(p),\hat{\bf H}(p) \rangle} \ .
    \label{eq:hermitdist}
\end{equation}
Geodesics are curves between two points on the space that minimize this arc-length.

From Eq.~(\ref{eq:hermittrans}) it follows that any $S$-matrix trajectory can be built up from continuous multiplication of the
identity matrix at $p = 0$ by $\hat {\bf 1}+ i \hspace{.06cm} dp \hspace{.06cm} \hat{\bf{H}}(p)$. In the limit that $dp \rightarrow
0$ the $S$-matrix trajectory is given by,
\begin{equation}
    \begin{split}
      \hat{\bf{S}}(p) = \lim_{dp\to 0} \hat{\bf{S}}(0) \prod_{n = 1}^{n = \frac{p}{dp}} \left( \hat {\bf 1} + i \hspace{.06cm} dp \hspace{.06cm} \hat{\bf{H}}((n - 1) \hspace{.06cm} dp ) \right)
    \;=\; \hat {\bf 1}\,{\cal P} \{ e^{i \int_0^p dp' \hat{\bf{H}}(p')} \} \ ,
    \end{split}
    \label{eq:momentaorder}
\end{equation}
where ${\cal P}$ acts as the momentum-ordering operator. Here the unitary $S$-matrix
is analogous to the unitary time-evolution operator. The initial $S$-matrix is the fixed point $\hat{\bf{S}}(p = 0)$ and the momentum evolution
propagates the system to another fixed point via the $\hat{\bf{H}}$ operator. It is straightforward to obtain
\begin{equation}
    \hat{\bf{H}}(p) = {1\over 2}\left( \dot{\phi} + \dot{\theta} \right)
 \hat   {\bf 1}
\ +\
{1\over 2}\left(\dot{\phi} - \dot{\theta} \right)
\left(\hat {\bf 1}+ \hat  {\bm \sigma} \cdot   \hat  {\bm \sigma}\right)/2 \ 
\label{eq:hermitdefined}
\end{equation}
where the dot denotes differentiation with respect to $p$.

\subsection{Entanglement power as a distance measure}
\label{sec:level1}

\noindent The HS distance also serves to obtain an operator definition and an alternate understanding of
the EP. Recall that the S-matrix is non-entangling when either $\phi = \theta $ or $\phi = \theta \pm
\pi$ \footnote{When $\phi = \theta \pm \pi$ the $S$-matrix acts as a
  swap gate on the incoming nucleon-nucleon state up to an overall
  phase. Likewise, when $\phi = \theta \pm \frac{\pi}{2}$, the
  $S$-matrix acts as a root-swap gate on the incoming nucleon-nucleon state
  up to an overall phase. }. Therefore, the non-entangling
$S$-matrices form a codimension-one subspace within the space of all
possible $S$-matrices. The EP of a given S-matrix, $\hat{\bf
  S}(\phi,\theta)$, is found to be,
\begin{equation}
    \mathcal{E}(\hat{\bf{S}}) = D(\hat{\bf S}(\phi,\theta), \hat{\bf{S}}(\theta,\theta))^2 \hspace{.1cm} D(\hat{\bf S}(\phi,\theta), \hat{\bf{S}}(\theta - \pi,\theta))^2 = N_p \sin^2{(\phi - \theta)} \ ,
    \label{eq:hsep}
\end{equation}
where the freedom in defining the HS norm has been used to set the normalization to $N_p$.
As both $\hat{\bf{S}}(\theta,\theta)$ and $\hat{\bf{S}}(\theta -
\pi,\theta)$ are non-entangling, the EP can be interpreted as a
measure of the distance from a given $S$-matrix to the two
non-entangling subspaces.  Using the HS distance
highlights the fact that the EP of an operator is a
state-independent measure of entanglement. 

\subsection{Embedding in $\mathbb{R}^4$}

\noindent While the HS distance provides a metric on the space of $S$-matrices, it is convenient to 
view the two-dimensional space on which the $S$-matrix propagates as a geometric embedding
in a higher-dimensional space. Recall that in the chosen isotropic coordinates, the first unitarity
constraint determines a three-sphere of fixed radius ($r=1$):
\begin{eqnarray}
 x^2\;+\; y^2\;+\; z^2\;+\; w^2 \;=\; r^2 \ .
   \label{eq:S3defgen2}
\end{eqnarray}
The isometry group of the three-sphere, $S^3$, is $SO(4)$ which is also the isometry group of $\mathbb{R}^4$.  The six $SO(4)$ generators can be constructed by considering the rotations
in the six planes that can be formed from the four Cartesian coordinates. The second unitarity constraint, Eq.~(\ref{eq:S3const}),
can be expressed in the two equivalent forms
\begin{eqnarray}
 (x\pm z)^2\;+\; (y\pm w)^2 \ =\ r^2 \ .
   \label{eq:constdefgen}
\end{eqnarray}
This leaves invariant two independent $SO(2)$ transformations. In addition, there are six discrete $\mathbb{Z}_2$ symmetries. Therefore, the isometry group of the
two-dimensional space on which the $S$-matrix propagates is
\begin{eqnarray}
SO(2)\otimes SO(2)\otimes \mathbb{Z}_2^6 \ .
   \label{eq:isomgroupuni}
\end{eqnarray}
These symmetries will be given explicitly in the next subsection.

As an embedding in $\mathbb{R}^4$, with metric 
\begin{eqnarray}\;
ds^2 \ =\ dx^2\;+\;dy^2\;+\;dz^2\;+\;dw^2 \ ,
   \label{eq:R4metricgen}
\end{eqnarray}
and with coordinate choice give in Eq.~(\ref{eq:PS}), one finds the flat two-dimensional Euclidean metric
\begin{eqnarray}\;
ds^2 \ =\ \oneht\;\left( d\phi^2\;+\;  d\theta^2\right)\ ,
   \label{eq:T2inR4}
\end{eqnarray}
with $\phi\in [0,2\pi]$ and $\theta\in [0,2\pi]$.
This metric describes the flat torus $\mathbb{T}^2\sim S^1 \otimes S^1\in\mathbb{R}^4$, where $S^1$
is the circle with isometry group $SO(2)$.\footnote{The flat torus also shows up in the construction of the Hilbert space of two qubits. See, for example, Ref.~\cite{BENGTSSON_2002}.} 

\subsection{Flat torus isometry group}

\noindent Here the action of the isometry group on the variables
$\phi$ and $\theta$ will be given explicitly. As $SO(4)\sim
SU(2)\otimes SU(2)$, the generators of $SO(4)$ may be given by the
generators of the two $SU(2)$'s, say $X_i$ and $Y_j$ with
$i,j=1,2,3$. The two $SO(2)$ isometries of the flat torus are then
generated by $X_3$ ($\equiv SO(2)_-$) and $Y_1$ ($\equiv SO(2)_+$),
where $X_3$ is a simultaneous rotation in the $x-y$ and $z-w$ plane by
the same amount, while $Y_1$ is a simultaneous rotation in the $x-w$
and $y-z$ plane by opposite amounts. The action of $X_3$ and $Y_1$ on
the angles $\phi$ and $\theta$ are given in Table~\ref{tab:cont}.
\begin{table}[h!]
\centering
  \begin{tabular}{ |c|c|c|c| } 
 \hline
$SO(2)_-$& $X_3$ & $\phi \mapsto \phi + \epsilon$ & $\theta \mapsto \theta + \epsilon$\\ 
$SO(2)_+$& $Y_1$ & $\phi \mapsto \phi + \epsilon$ & $\theta \mapsto \theta - \epsilon$\\ 
 \hline
  \end{tabular}
  \caption{Continuous isometries of the flat torus.}
 \label{tab:cont}
\end{table}
Note that these are simply the translational symmetries that one would expect on a flat manifold.
The action of the six $\mathbb{Z}_2$ symmetries on the angles $\phi$ and $\theta$ are given in Table~\ref{tab:disc}.
\begin{table}[h!]
\centering
\begin{tabular}{ |c|c|c|c| } 
 \hline
a& $\text{- + - +}$ & $\phi \mapsto \pi - \phi$ & $\theta \mapsto \pi - \theta$\\ 
b& $\text{+ - + -}$ & $\phi \mapsto - \phi$ & $\theta \mapsto - \theta$\\ 
c& $\text{+ + - -}$ & $\phi \mapsto \theta $ & $\theta \mapsto \phi$\\ 
d& $\text{+ - - +}$ & $\phi \mapsto - \theta$ & $\theta \mapsto -\phi$\\
e& $\text{- + + -}$ & $\phi \mapsto \pi - \theta$ & $\theta \mapsto \pi - \phi$\\
f& $\text{- - + +}$ & $\phi \mapsto \pi + \theta$ & $\theta \mapsto \pi + \phi$ \\
 \hline
\end{tabular}
  \caption{Discrete isometries of the flat torus. Here $(+-+-)$ corresponds to $(x,y,z,w)\mapsto (x,-y,z,-w)$, {\it etc}. Note that the signs of all values of $\pi$ are arbitrary.}
 \label{tab:disc}
\end{table}
It is clear that the  UV/IR conformal symmetries of the LO in the ERE and in the conformal range model are
in correspondence with discrete flat-torus isometries. This will be important in what follows.

\section{$S$-matrix theory of scattering}
\label{sec:smattheory}

\subsection{Action principle}

\noindent The s-wave nucleon-nucleon $S$-matrix is a one-dimensional trajectory that lives on a flat torus, which is a
two-dimensional Euclidean space with periodic boundary conditions on the two phase-shift coordinates.
Straight line trajectories on this flat space are geodesics, which
are formally obtained by minimizing an action which represents a path in the space. In general, 
$S$-matrix trajectories will not be geodesics and therefore external forces must be present.
The action for a general parameterization\footnote{Note that this form avoids the square root in the Lagrangian while allowing inaffine parameterizations.} of a curve on a space with metric tensor $g_{ab}$
can be taken as~\cite{garay2019classical}
 \begin{equation}
 \int L\left(\mathcal{X},\dot{\mathcal{X}} \right)d\sigma \ =\ \int \left({\bf{N}}^{-2}g_{ab} \dot{\mathcal{X}}^a \dot{\mathcal{X}}^b \ -\  \mathbb{V(\mathcal{X})} \right){\bf{N}}d\sigma
    \label{eq:action}
 \end{equation}
 where $\sigma$ is the parameter (affine or inaffine),
 $\dot{\mathcal{X}}\equiv{d\mathcal{X}}/{d\sigma}$, and
 $\mathbb{V(\mathcal{X})}$ is an external potential which is assumed
 to be a function of $\mathcal{X}$ only\footnote{This assumption will have to be relaxed when considering projections onto parts of a space.}. Minimizing the action or
 equivalently solving the Euler-Lagrange equations gives the  trajectory equation
 \begin{equation}
\ddot{\mathcal{X}}^a \ +\ {}_g\Gamma^a_{\ bc} \dot{\mathcal{X}}^b \dot{\mathcal{X}}^c \ =\ \kappa(\sigma) \dot{\mathcal{X}}^a
   \ -\ \oneht {\bf{N}}^2 g^{ab}  \partial_b\mathbb{V(\mathcal{X})} \ ,
    \label{eq:exEOMfromact}
 \end{equation}
 where ${}_g\Gamma^a_{\ bc}$ are the Christoffel symbols for the metric $g_{ab}$, and
 \begin{equation}
\kappa(\sigma) \ \equiv \ \frac{\dot{\bf{N}}}{\bf{N}} = \frac{d}{d\sigma}\ln \frac{d\lambda}{d\sigma} \ .
    \label{eq:inaffinity}
 \end{equation}
Here $\kappa$ is the inaffinity~\cite{blau2020}, which vanishes when $\sigma=\lambda$ with $\lambda$ an affine parameter. For constant potential, the trajectory equation reduces to the geodesic equation.

\subsection{Geodesics on the flat torus}

\noindent With $\mathcal{X}^1=\phi$ and $\mathcal{X}^2=\theta$ and omitting the external force term, 
the equations for geodesics are then read off to be:
\begin{eqnarray}
&&\ddot{\phi} \  =\   \kappa(\sigma)\dot{\phi} \ \ , \nn \\
&&\ddot{\theta}  \  = \  \kappa(\sigma)\dot{\theta}  \ .
     \label{eq:R4geoaffine}
\end{eqnarray}
In the affine case, $\sigma=\lambda$ and $\kappa=0$, $\phi$ and $\theta$ are linear
functions of $\lambda$, and the most general solution is a
straight line in the Euclidean ($\phi$-$\theta$) plane. In the
non-affine case $\phi$ and $\theta$ can be arbitrary functions of
$\sigma$, and the most general solution is again a straight line in
the Euclidean ($\phi$-$\theta$) plane. For instance, say $\phi=f(\sigma)$ and
$\theta=g(\sigma)$ with $f$ and $g$ arbitrary differentiable functions.
Then,
\begin{eqnarray}
\frac{\ddot{\phi}}{\dot{\phi}} \  =\  \frac{\ddot{\theta}}{\dot{\theta}} \  =\   \kappa(\sigma) 
     \label{eq:R4geoaffine2}
\end{eqnarray}
implies $f=c_1+c_2\;g$, giving the desired result. In this way, using the
inaffinity, the effective range expansion can be built up to any order.
For instance, to leading order, assuming that $\phi=\theta$ and $\sigma=p$, one can choose
$\phi=\theta=-2\tan^{-1}\!\left( a p\right)$  with
\begin{eqnarray}
  \kappa(p) \ =\ -\frac{2 a^2 p}{1+a^2 p^2} \ .
 \label{eq:baseinaffinity}
\end{eqnarray}
At NLO in the ERE one can choose $\phi=\theta=2\cot^{-1}\!\left( -\frac{1}{ap} + \oneht r p\right)$ where $r$ is the effective range,
with a corresponding $\kappa$, and so the momentum dependence of the full phase shift may be built up by appropriate
choice of inaffinity.

\subsection{Entanglement forces and the ERE}

\begin{figure}[!ht]
\centering
\includegraphics[width = 0.45\textwidth]{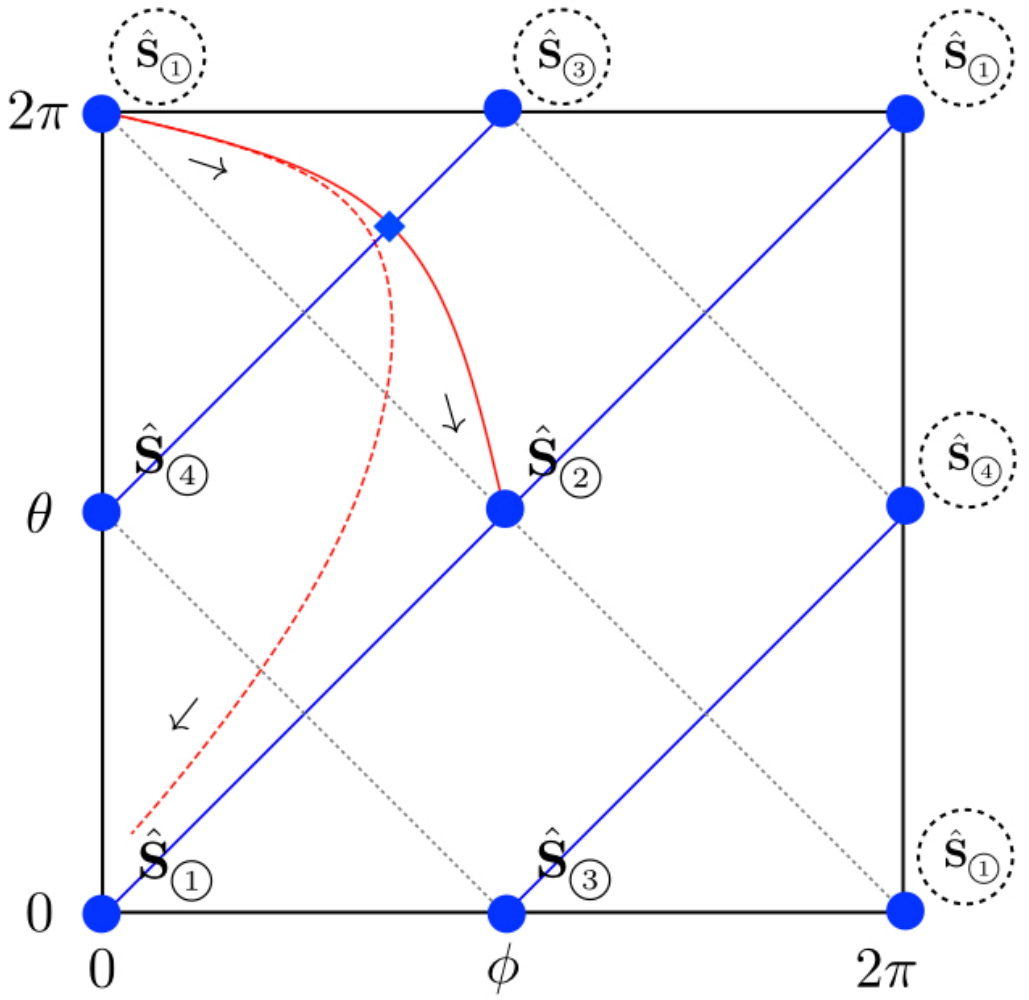}
\caption{The $S$-matrix embedding in $\mathbb{R}^4$ as the flat torus. The blue dots are the RG fixed points of the $S$-matrix and the image
  fixed points are within dotted circles. All diagonal lines are geodesics. The solid blue lines are geodesics with vanishing entanglement,
  and the dotted black lines are geodesics with non-vanishing entanglement. The blue diamond is the fixed point of the UV/IR conformal symmetry.
  The physical $S$-matrix taken from the Nijmegen phase shift analysis~\cite{NNOnline} (PWA93) is the dashed red curve, and LO in the ERE is the
  solid red curve.}
    \label{fig:FT}
\end{figure}

\noindent Recall that the EP is
\begin{eqnarray}
{\mathcal E}({\hat {\bf S}})
& = &  {N}_{\cal P}\;  \sin^2\left(\phi-\theta\right) \ .
  \label{eq:R31ep2}
\end{eqnarray}
The EP preserves all of the discrete isometries of the flat torus, however
the $SO(2)_+$ translational symmetry is broken; only translations in one
direction on the flat torus preserve spin-entangling effects. The blue lines
in Fig.~(\ref{fig:FT}) are lines of vanishing entanglement and all lines parallel are lines of equi-entanglement which are protected by the 
$SO(2)_-$ invariance.

Therefore in order to find a non-geodesic solution with $\phi\neq \theta +n\pi$, an external entangling force must be
present. The general equation, Eq.~(\ref{eq:exEOMfromact}) gives
\begin{eqnarray}
&&\ddot{\phi} \ =\  \kappa(\sigma)\dot{\phi}  \ -\ {\bf{N}}^2\,\partial_\phi\mathbb{V}(\phi,\theta)   \ \ , \nn \\
&&\ddot{\theta}\ =\  \kappa(\sigma)\dot{\theta} \ -\  {\bf{N}}^2\,\partial_\theta\mathbb{V}(\phi,\theta) \ .
     \label{eq:SQ1gesNAR4}
\end{eqnarray}
If a solution for $\phi$ and $\theta$ is specified, these two coupled equations have three unknowns given by
the inaffinity and the components of the force in the two directions in the plane.

\subsection*{LO in the ERE}

\noindent Consider LO in the ERE. With $\sigma =p$, the solution is once again
\begin{eqnarray}
\phi & =& -2\tan^{-1}\!\left( a_0 p\right) \ \ \ , \ \ \ \theta\ =\ -2\tan^{-1}\!\left( a_1 p\right) \ ,
 \label{eq:LOinaffineSOL2}
\end{eqnarray}
and is exhibited in Fig.~(\ref{fig:FT}). (The physical trajectory with the Nijmegen phase shift analysis~\cite{NNOnline} (PWA93) is also shown.)
The conformal transformation $p\mapsto (|a_1 a_0| p)^{-1}$ is in
correspondence with the isometry $\mathbb{Z}^e_2$ for $a_1 a_0 >0$ (doubly-(un)bound) and
$\mathbb{Z}^f_2$ for $a_1 a_0 <0$ (singly-bound). These isometries leave the angle
$\phi+\epsilon\,\theta$ invariant with $\epsilon\,=-1$ for $a_1 a_0
>0$ and $\epsilon\,=+1$ for $a_1 a_0 <0$. For these isometries, $\mathbb{V}(\phi + \epsilon \theta, \phi - \epsilon \theta) = \mathbb{V}(\phi + \epsilon \theta, -(\phi - \epsilon \theta))$ and the simplest solution is  $\mathbb{V}(\phi,\theta)=\mathbb{V}(\phi+\epsilon\,\theta)$. This implies
$\partial_\phi\mathbb{V}=\epsilon\,\partial_\theta\mathbb{V}$, which renders the system integrable: the equations decouple to
\begin{eqnarray}
&&\ddot{\phi}+\epsilon\,\ddot{\theta} \ =\   \kappa \left(  \dot{\phi}+\epsilon\,\dot{\theta}  \right)  \ +\ 2\,\mathbb{F} \ \ , \nn \\
&&\ddot{\phi}-\epsilon\,\ddot{\theta} \ =\   \kappa \left(  \dot{\phi}-\epsilon\,\dot{\theta}  \right)  \ ,
\end{eqnarray}
where the external force is given by $\mathbb{F} \equiv -{\bf{N}}^2\,\partial_\phi\mathbb{V}$.

The inaffinity and the force are determined algebraically to be
\begin{eqnarray}
  &&{\bf{N}} \ = \ c_1\left(\dot{\phi} -\epsilon\,\dot{\theta} \right)  \ \ ,  \ \
  \kappa \ = \ \left(\frac{\ddot{\phi} -\epsilon\,\ddot{\theta}}{\dot{\phi} -\epsilon\,\dot{\theta}}\right)  \ \ , \nn \\
  &&\qquad\quad\mathbb{F}\ = \ -\epsilon \left(\frac{\ddot{\phi}\,\dot{\theta} -\ddot{\theta}\,\dot{\phi}}{\dot{\phi} -\epsilon\,\dot{\theta}}\right) \  ,
     \label{eq:SQ1gesNAR4sol}
\end{eqnarray}
where $c_1>0$ is an integration constant. In terms of the scattering lengths, one finds
\begin{eqnarray}
&&\hspace{-.5in} {\bf{N}} \ = \ -\frac{2 c_1 \left(a_0 -\epsilon\,a_1 \right)\left(1- \epsilon\,a_0a_1 p^2\right)}{\left(1+a_0^2 p^2\right)\left(1+a_1^2 p^2\right)} 
  \ \ , \ \ \kappa(p) \ =\ -\frac{2 a_0^2 p}{1+a_0^2 p^2} -\frac{2 a_1^2 p}{1+a_1^2 p^2} -\frac{2 a_0a_1 p}{\epsilon-a_0a_1 p^2}\ ,
\label{eq:R4solLOERT} 
\end{eqnarray}
and the force is
\begin{eqnarray}
\mathbb{F} & = & -  \frac{4\epsilon\, a_0 a_1 p \left( a^2_1-a_0^2\right)}
    {\left(1+a_0^2 p^2\right)\left(1+a_1^2 p^2\right) \big\lbrack a_1\left(1+a_0^2p^2\right)\epsilon-a_0\left(1+a_1^2p^2\right)\big\rbrack} \ .
\label{eq:R4solLOERTF} 
\end{eqnarray}
The inaffinity and the entangling force are complicated non-local functions of the momentum; this is expected as locality plays no role in
constraining the geometric description.
\begin{figure}[!ht]
\centering
\includegraphics[width = 0.85\textwidth]{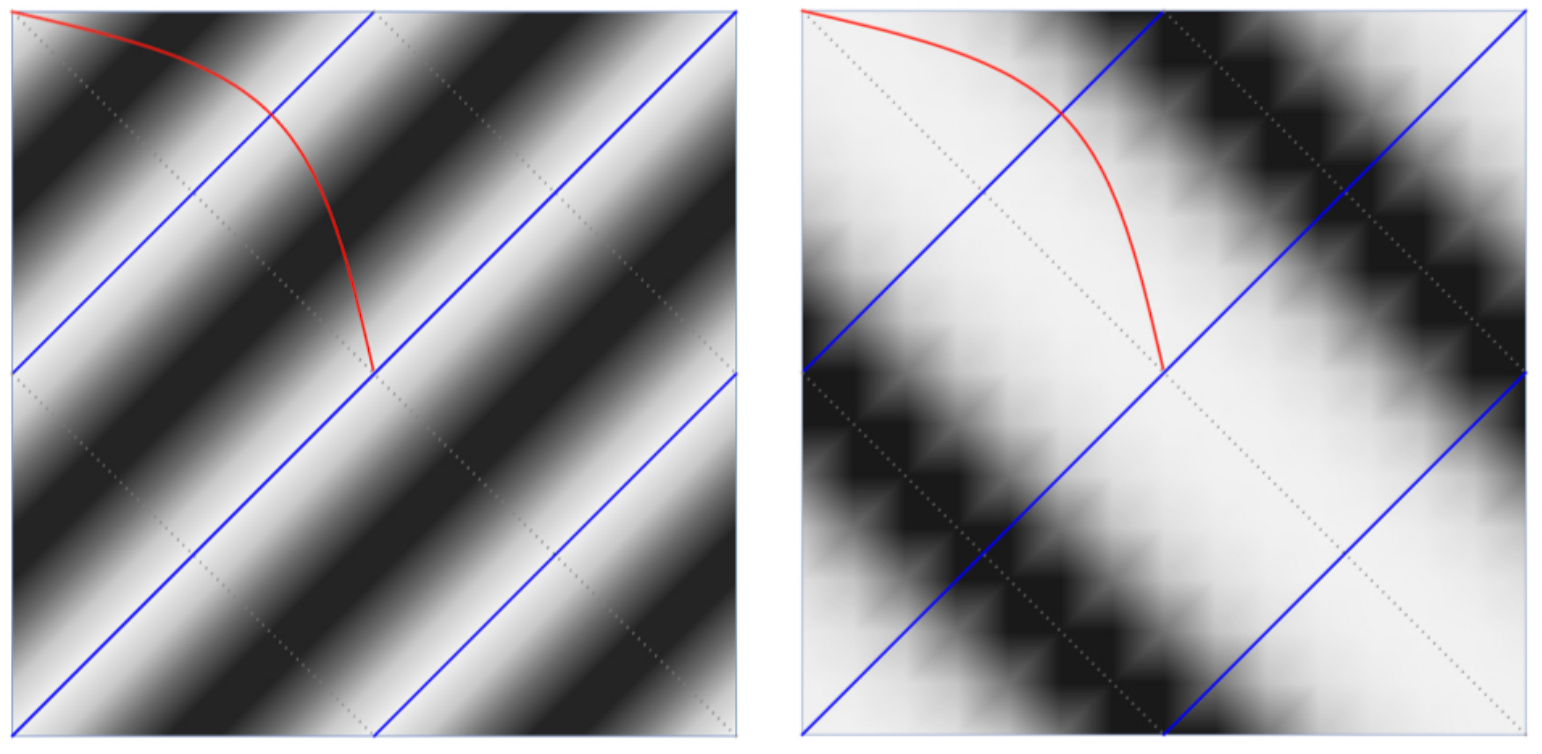} 
\caption{Heat maps on the flat torus illustrating regions of equi-entanglement and equi-potential. The red curve in both panels is the $S$-matrix trajectory at LO in the ERE. The EP of the $S$-matrix is shown in the left panel. Lighter shade indicates smaller EP.
The external potential given by Eq.~(\ref{eq:entPOT}) is shown in the right panel. Lighter shade indicates smaller potential.}
    \label{fig:heatmaps}
\end{figure}

Integrating the force gives the external potential,
\begin{eqnarray}
\mathbb{V}(\phi,\theta) & = &  c_2\, \tan^2\left(\oneht(\phi+\epsilon\,\theta)\right) \ ,
\label{eq:entPOT} 
\end{eqnarray}
where
\begin{eqnarray}
  c_2  & = &   \frac{|a_0 a_1|}{\left(|a_0|+|a_1| \right)^2 c_1^2} \ .
\label{eq:c2inaffine} 
\end{eqnarray}
With $c_1=1$, $c_2$ is a dimensionless coupling constant that ranges from $0$ to $0.25$. In the physical case $c_2=0.152$.
The potential is constrained by the assumed discrete symmetries and, defined over the entire manifold, is 
invariant with respect to the $SO(2)_\epsilon$ translational
symmetry. Therefore, only in the doubly-(un)bound case ($\epsilon=-1$),
does the potential share the symmetry of the entanglement power. This
asymmetry is a fundamental feature of the geometric description. The
relative sign of the scattering lengths picks one preferred direction
on the flat torus. However, the entanglement power universally breaks
the  $SO(2)_+$ isometry of the flat torus.  Heat maps for the entanglement power
and the external potential, $\mathbb{V}(\phi,\theta)$, are shown for
the physical case in Fig.~(\ref{fig:heatmaps}). It is worth
re-considering the plot of the EP given in Fig.~(\ref{fig:EPcurves})
using the heat map as a guide; the $S$-matrix trajectory leaves the
initial fixed point, ascends to a maximum of the EP, then descends to
the zero of the EP at the UV/IR conformal fixed point which lies on
the blue geodesic, and then rises again to the maximum before hitting the
final fixed point. 

\begin{figure}[!ht]
\centering
\includegraphics[width = 0.4\textwidth]{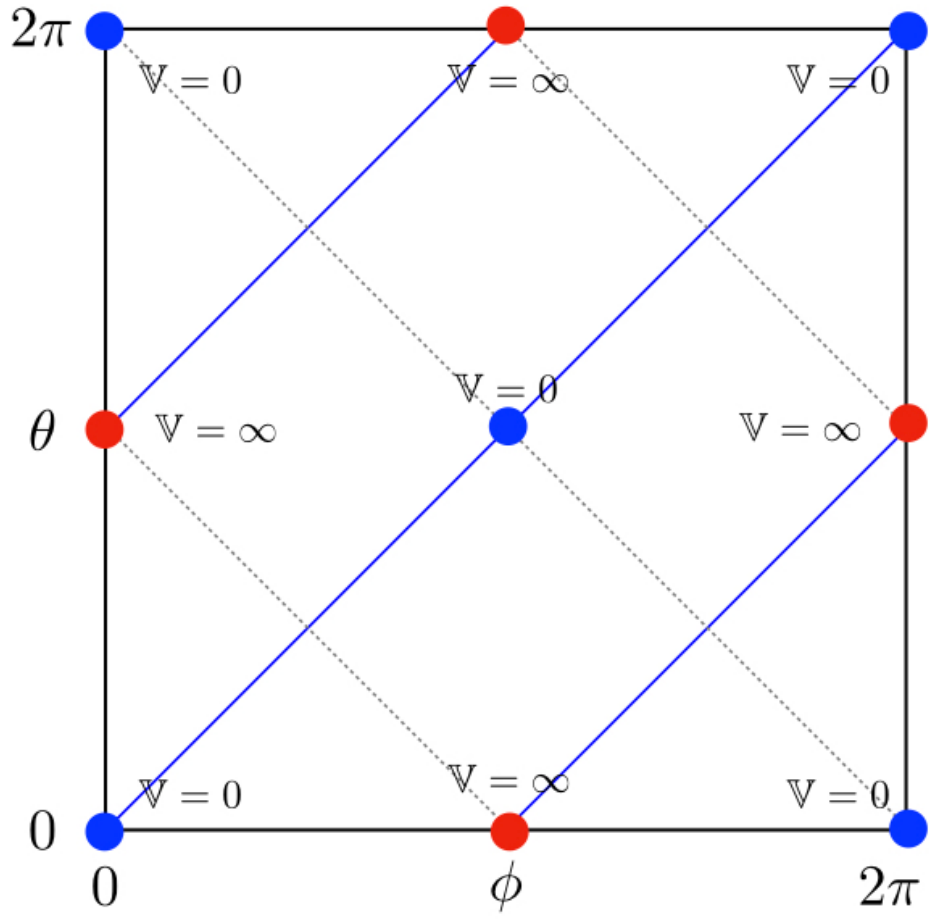}
\caption{Discrete values of the singly-bound entangling potential on the flat torus for finite values of the scattering lengths. The potential vanishes at the blue dot fixed points and is infinite at the red dot
  fixed points.}
    \label{fig:FTFR}
\end{figure}

It is straightforward to understand why the potential takes the simple
harmonic form of Eq.~(\ref{eq:entPOT}). It follows from the solution
Eq.~(\ref{eq:LOinaffineSOL2}), that for any finite values of the
scattering lengths, only the $\hat {\bf S}_{\tiny{\circled{1}}}$ and
$\hat {\bf S}_{\tiny{\circled{2}}}$ fixed points (and their images)
can be accessed by an $S$-matrix trajectory. Therefore the potential
at these fixed points should be finite. And because of the symmetry
implied by the condition that $\phi+\epsilon\,\theta$ be invariant,
the potential must take the same constant value at each fixed
point. As the $\hat {\bf S}_{\tiny{\circled{3}}}$ and $\hat {\bf
  S}_{\tiny{\circled{4}}}$ fixed points (and their images) cannot be
accessed by an $S$-matrix trajectory, the potential should be infinite
at these fixed points. (See Fig.~(\ref{fig:FTFR}).) On the flat torus,
the potential must be a harmonic function and therefore these
constraints imply that $\mathbb{V}$ is proportional to
$\sec(\oneht(\phi+\epsilon\,\theta))$ to some even power. Shifting the
potential by a constant to give vanishing potential at the accessible
fixed points then gives Eq.~(\ref{eq:entPOT}). If one of the channels
approaches unitarity, then the potential changes form; for instance, if
$a_0$ is taken to negative infinity and $a_1$ is held fixed at its
physical value, then the potential vanishes and the $S$-matrix trajectory 
begins at fixed point
$\hat {\bf S}_{\tiny{\circled{3}}}$ and moves to $\hat {\bf
  S}_{\tiny{\circled{2}}}$ along a geodesic.

\begin{figure}[!ht]
\centering
\includegraphics[width = 0.65\textwidth]{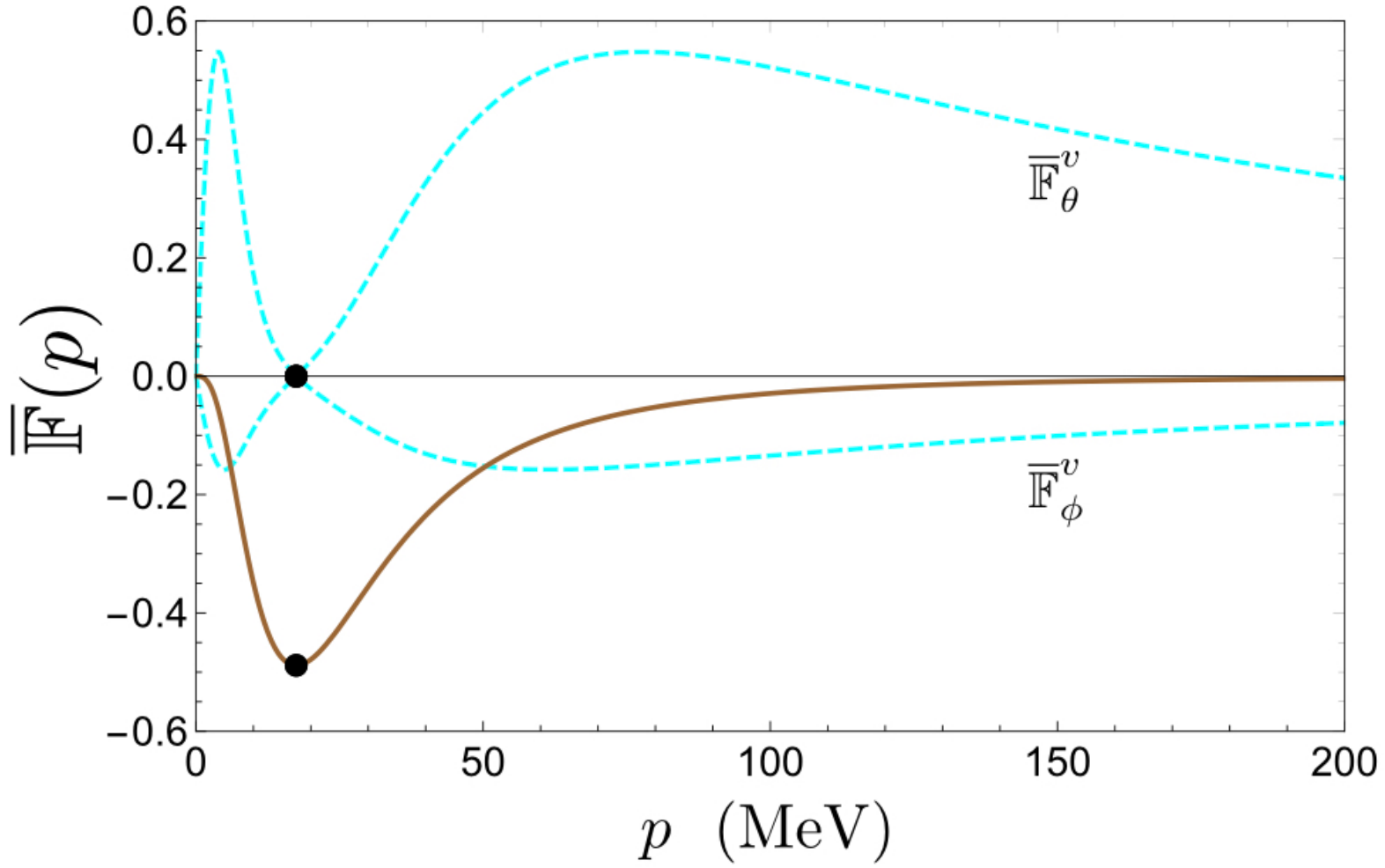}
\caption{The dimensionless entangling force plotted versus momentum in the physical case (solid brown line). The black dot corresponds to the fixed point of the conformal transformation where the $S$-matrix
crosses the geodesic and the EP vanishes. Also plotted are the viscous inaffinity forces (dashed cyan lines).}
    \label{fig:conF}
\end{figure}

It is convenient to define the conformal derivative and coordinates
\begin{eqnarray}
D \equiv p\frac{d}{dp} \ \ , \ \ {\overline{\mathbb{F}}}\equiv p^2\, \mathbb{F} \ \ , \ \   {\overline{\mathbb{F}}_a^v}  \equiv \left( 1\;+\;p\, {\kappa}\right)D{\cal X}_a   \ .
  \label{eq:moebiusconfderivdef}
\end{eqnarray}
The trajectory equations are then
\begin{eqnarray}
&&D^2 \phi \ =\  {\overline{\mathbb{F}}_\phi^v}\,+\, {\overline{\mathbb{F}}}   \ , \nn \\
&&D^2 \theta \ =\ {\overline{\mathbb{F}}_\phi^v}\,+\, \epsilon\;{\overline{\mathbb{F}}}  \ ,
     \label{eq:SQ1gesNAR4go2}
\end{eqnarray}
which have the manifest conformal invariance
\begin{eqnarray}
D \to -D \ \ , \ \ \phi \leftrightarrow \epsilon\theta \ \ , \ \  {\overline{\mathbb{F}}}\to {\overline{\mathbb{F}}} \ \ , \ \  {\overline{\mathbb{F}}_\phi^v} \leftrightarrow \epsilon{\overline{\mathbb{F}}_\theta^v}  \ .
\end{eqnarray}
The dimensionless entangling forces in the physical case are plotted in Fig.~(\ref{fig:conF}). While the entangling force ${\overline{\mathbb{F}}}/{\bf{N}}^2$ is physical, the viscous inaffinity forces
${\overline{\mathbb{F}}_\phi^v}$ and ${\overline{\mathbb{F}}_\theta^v}$ can be altered or removed by reparameterization. However, it is critical to note that these viscous forces build the momentum
variable which is the experimental knob which allows the measurement of the $S$-matrix.

It is straightforward to find the solution corresponding to vanishing inaffinity, described by an affine parameter $\lambda$ that is unrelated to any experimental knob. In this case, the trajectory is governed by the simple Lagrangian,
\begin{eqnarray}
L &=& \oneht \left( \dot{\phi}\,+\,\dot{\theta} \right) \ -\  c_2\, \tan^2\left(\oneht(\phi+\epsilon\,\theta)\right) \ ,
\label{eq:LagAFFLO}
\end{eqnarray}
where the dot denotes differentiation with respect to $\lambda$. Variation of the Lagrangian gives rise to the
geodesic equations
\begin{eqnarray}
&&\ddot{\phi}+\epsilon\,\ddot{\theta} \ =\   -\ 2\,c_2\, \sec^2\left(\oneht(\phi+\epsilon\,\theta)\right) \tan\left(\oneht(\phi+\epsilon\,\theta)\right) \ , \nn \\
&&\ddot{\phi}-\epsilon\,\ddot{\theta} \ =\   0\ .
\end{eqnarray}
The affine parameter can be chosen to give the solution
\begin{eqnarray}
&&{\phi}+\epsilon\,{\theta} \ =\   \ 2\,\sin^{-1}\left(\sqrt{1-4c_2}\,\sin{\frac{\lambda}{2}}\right) \ , \nn \\
&&{\phi}-\epsilon\,{\theta} \ =\   \lambda \ .
\end{eqnarray}
The $S$-matrix trajectory from fixed point $\hat {\bf S}_{\tiny{\circled{1}}}$ to fixed point $\hat {\bf S}_{\tiny{\circled{2}}}$ is then
found to be
\begin{eqnarray}
&&{\phi}(\lambda ) \ =\   \frac{1}{2} \left( 2\sgn(|a_0|-|a_1|)\sin^{-1}\left(\Big|\frac{a_0+\epsilon a_1}{a_0-\epsilon a_1}\Big|\sin{\frac{\lambda}{2}}\right)+\lambda \right) \ , \nn \\
&&{\theta}(\lambda ) \ =\    \frac{\epsilon}{2} \left( 2\sgn(|a_0|-|a_1|)\sin^{-1}\left(\Big|\frac{a_0+\epsilon a_1}{a_0-\epsilon a_1}\Big|\sin{\frac{\lambda}{2}}\right)-\lambda \right) \ ,
\end{eqnarray}
where $c_2$ has been fixed to match the inaffine solution at the conformal fixed point (and is consistent with Eq.~(\ref{eq:c2inaffine}) with $c_1=1$).
This $S$-matrix trajectory is exactly equivalent to LO in the ERE.

\subsection*{Conformal range model}

\begin{figure}[!ht]
\centering
\includegraphics[width = 0.5\textwidth]{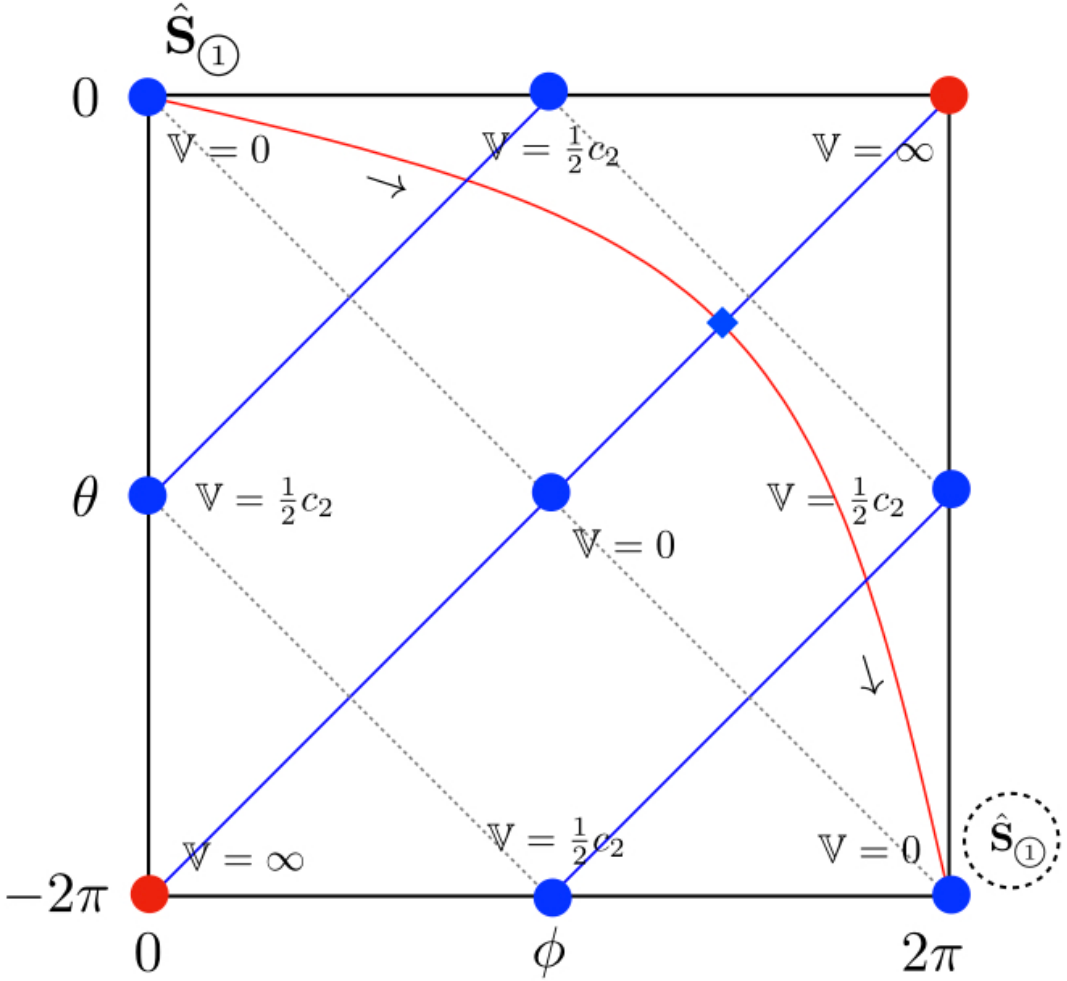}
\caption{The $S$-matrix trajectory in the conformal range model on the
  flat torus. Physical values of the scattering lengths and
  $\lambda=1/4$ have been used. The blue diamond is the fixed point of
  the UV/IR conformal symmetry. The blue dots are the fixed points of
  the $S$-matrix and the image fixed points are within dotted
  circles. All diagonal lines are geodesics. The solid blue lines are
  geodesics with vanishing entanglement, and the dotted black lines
  are geodesics with non-vanishing entanglement.}
    \label{fig:FTmod2}
\end{figure}

\noindent Consider now the solution of the conformal range model
\begin{eqnarray}
\phi & =& -2\tan^{-1}\!\left( \frac{a_0 p}{1-\lambda a_0^2 p^2} \right) \ \ \ , \ \ \ \theta\ =\ -2\tan^{-1}\!\left( \frac{a_1 p}{1-\lambda a_1^2 p^2} \right) \ ,
 \label{eq:confRmodSOL}
\end{eqnarray}
which is exhibited on the flat torus in Fig.~(\ref{fig:FTmod2}) in the case of physical
scattering lengths and $\lambda=1/4$.  The conformal transformation
$p\mapsto (\lambda |a_1 a_0| p)^{-1}$ is in correspondence with the
isometry $\mathbb{Z}^d_2$ for $a_1 a_0 >0$ and $\mathbb{Z}^c_2$ for
$a_1 a_0 <0$. These isometries again leave the angle
$\phi+\epsilon\,\theta$ invariant with $\epsilon\,=-1$ for $a_1 a_0
>0$ and $\epsilon\,=+1$ for $a_1 a_0 <0$. Therefore the algebraic form
of the solution is identical to what we found in the LO ERE,
i.e. Eq.~(\ref{eq:SQ1gesNAR4sol}).  For arbitrary values of $\lambda$
the entangling potential is cumbersome. However, in the special case
$\lambda=1/4$ the potential reduces to
\begin{eqnarray}
\mathbb{V}(\phi,\theta) & = &  \oneht c_2\, \tan^2\left(\onefourth(\phi+\epsilon\,\theta)\right) \ .
\label{eq:entPOTmod2} 
\end{eqnarray}
It is clear from the solution of Eq.~(\ref{eq:confRmodSOL}) that only the $\hat {\bf S}_{\tiny{\circled{1}}}$ trivial fixed point and its image
can be reached by an $S$-matrix trajectory. Fig.~(\ref{fig:FTmod2}) indicates discrete values of the potential
on the flat torus. Note that this potential is periodic only by a shift of $\phi+\epsilon\,\theta$ by $4\pi$,
in contrast with the LO ERE potential.

\section{Inelasticity and holography}
\label{sec:inelatandholo}

\subsection{Inelasticity and the bulk}

\noindent So far it has been assumed that all inelastic thresholds in
the scattering process have been pushed to infinity so that the
$S$-matrix is unitary and defined for all (positive) center-of-mass
momenta $p$.  One might imagine controlling the inelasticity in
nucleon-nucleon scattering by varying the light-quark masses in QCD to
adjust the threshold for pion production. In the chiral limit, at
scattering threshold there will be pion radiation which, for present
purposes, is not measured and is removed from the system as a loss of
unitarity. In the $S$-matrix formalism, the inclusion of some generic
inelastic scattering process is achieved by replacing the
single-channel $S$-matrices by $\eta_0 \exp{2 i \delta_0}$ and $\eta_1
\exp{2 i \delta_1}$, with the inelasticity parameters satisfying
$0<\eta_{0,1}<1$. Assuming here that $\eta_{0}=\eta_{1}=r$,
inelasticity can be realized in the $\mathbb{R}^4$ embedding
coordinates of Eq.~(\ref{eq:PS}) through variation of $r$\footnote{Note that the assumption that the two channels couple to the same source
of inelasticity is critical; if $\eta_0$ and $\eta_1$ are allowed to vary independently then the resulting geometry is radically different.}.
The $S$-matrix with inelasticities present, $\hat {\bf S}_I$, then satisfies
the formal relation
\begin{eqnarray}
\hat {\bf S}_I^\dagger  \hat {\bf S}_I & = & \hat {\bf S}^\dagger  \hat {\bf S} \ -\ \sum_\gamma |\gamma\rangle \langle \gamma| \ ,
  \label{eq:SSdagINa}
\end{eqnarray}
where $\hat {\bf S}$ is the unitary $S$-matrix of the total system (nucleon-nucleon and inelastic channels), and $\gamma$ represents
an inelastic contribution. It then follows that
\begin{eqnarray}
\left( 1\;-\;r^2 \right)\;\hat {\bf 1}  &= & \sum_\gamma |\gamma\rangle \langle \gamma| \ .
  \label{eq:SSdagINb}
\end{eqnarray}
It is clear that, in general, $r$ depends in a complicated way on
momentum and involves a summation over all kinematically-allowed final
states. In the nucleon-nucleon system, the leading inelasticities
arise from pion production, which in principle is calculable using
chiral-perturbation theory methods. So for instance, the excluded
states may be of the form
\begin{eqnarray}
\sum_\gamma |\gamma\rangle \langle \gamma| & =& |NN\pi\rangle \langle NN\pi| \ +\ \ldots \ +\ |NN\pi\ldots\pi\rangle \langle NN\pi\ldots\pi| \ +\ \ldots \ .
  \label{eq:SSdagINc}
\end{eqnarray}
The momentum flow equation with inelasticities present is
\begin{eqnarray}
 p\frac{d}{dp} {\hat {\bf S}_I} & = &  r\; \left( p\frac{d}{dp} {\hat {\bf S}} \ +\ \left(p\frac{d}{dp}\ln r \right){\hat {\bf S}} \right) \ .
\end{eqnarray}
In addition to the four RG fixed points on the unitary boundary at $r=1$, there is now an additional fixed point at $r=0$.
All the fixed points have vanishing spin entanglement as the EP generalizes to
\begin{eqnarray}
{\mathcal E}({\hat {\bf S}_I})
& = &  {N}_{\cal P}\; r^4 \sin^2\left(\phi-\theta\right) \ ,
  \label{eq:R31ep}
\end{eqnarray}
which drops off rapidly when $r<1$.

\subsection{Embedding in $\mathbb{R}^4$}

\noindent With the three coordinates $\mathcal{X}^1=r$, $\mathcal{X}^2=\phi$, and $\mathcal{X}^3=\theta$ embedded in $\mathbb{R}^4$,
the metric of the bulk space takes the form
\begin{eqnarray}\;
ds^2 \ =\ dr^2 \ +\ \oneht\;r^2 \left( d\phi^2\;+\;  d\theta^2\right)\ ,
   \label{eq:T2inR4varyr}
\end{eqnarray}
with flat-torus boundary at $r=1$. The hyperbolic space described by this metric has scalar curvature
\begin{eqnarray}\;
R \ =\ -\frac{2}{r^2} \ .
   \label{eq:bulkcs}
\end{eqnarray}
There is therefore a singularity\footnote{The Kretschmann scalar is given by $K=R^{ijkl}R_{ijkl}=R^2=4/r^4$.} at the fixed point $r=0$ where there is total loss of unitarity and vanishing EP.
The Einstein tensor has one non-vanishing component given by
\begin{eqnarray}\;
G_{11} \ =\ \frac{1}{r^2} \ .
   \label{eq:Gbarr}
\end{eqnarray}
Therefore the bulk metric solves the Einstein equation\footnote{This
  solution is reminiscent of Vaidya metrics in spacetime, which are
  generalizations of the Schwarzschild metric with time-dependent mass
  function, and describe null dust forming a black hole~\cite{blau2020}.}
\begin{eqnarray}\;
G_{ij} \ =\ \frac{1}{r^2} \delta_i^1 \delta_j^1\ .
   \label{eq:EER4varr}
\end{eqnarray}
It is worth emphasizing again that this singular hyperbolic geometry results
as a consequence of the assumption that the inelasticity in the two
channels of scattering are correlated. If $\eta_0$ and $\eta_1$ are
taken to be independent, then the resulting embedding simply yields a
parametric representation of the flat four-dimensional Euclidean space, $\mathbb{R}^4$.

\subsection{Geodesics in the bulk}

\begin{figure}[!ht]
\centering
\includegraphics[width = 0.38\textwidth]{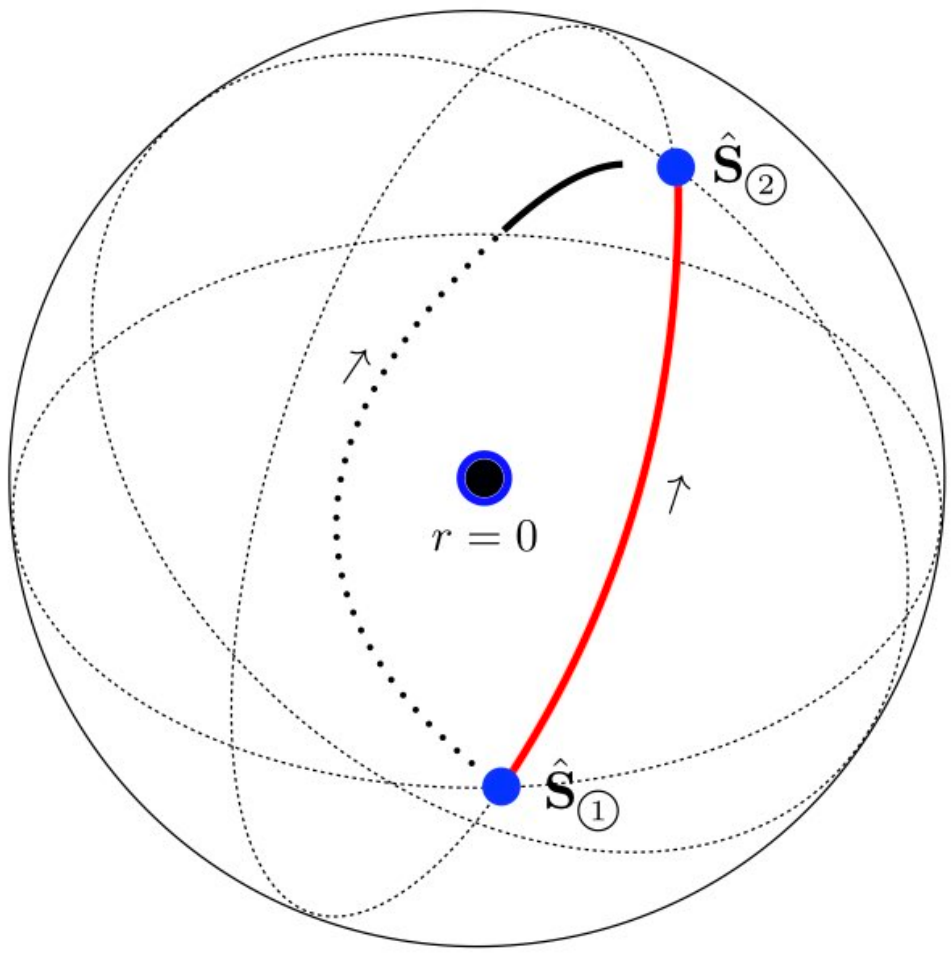}
\caption{Bulk space with flat torus on the boundary. The unitary $S$-matrix lives on the flat-torus boundary and here corresponds to the (solid red) trajectory
  between the RG fixed points. The bulk (dotted black) trajectory begins on the boundary and through inelastic loss enters the bulk, while avoiding the singularity,
  and then comes back out to the boundary without quite reaching the fixed point.}
  \label{fig:bulkdiag}
\end{figure}

\noindent An interesting question is whether the $S$-matrix
trajectories on the boundary can be recovered in the bulk space. That
is, is it possible to engineer a bulk potential which gives a bulk
trajectory which in turn reproduces the boundary trajectory
corresponding to LO in the ERE? (An illustration is provided in 
Fig.~(\ref{fig:bulkdiag})). Trajectories on the flat-torus
boundary are unitary, by construction, and are in correspondence with
local interactions in the EFT. By contrast, inelastic effects are in
correspondence with non-local interactions in the EFT. Geodesics in
both affine and inaffine parameterizations will be considered first,
as the construction of these solutions enables a straightforward
engineering of the potential which governs the trajectories in the LO
ERE solution.

\begin{figure}[!ht]
\centering
\includegraphics[width = 0.98\textwidth]{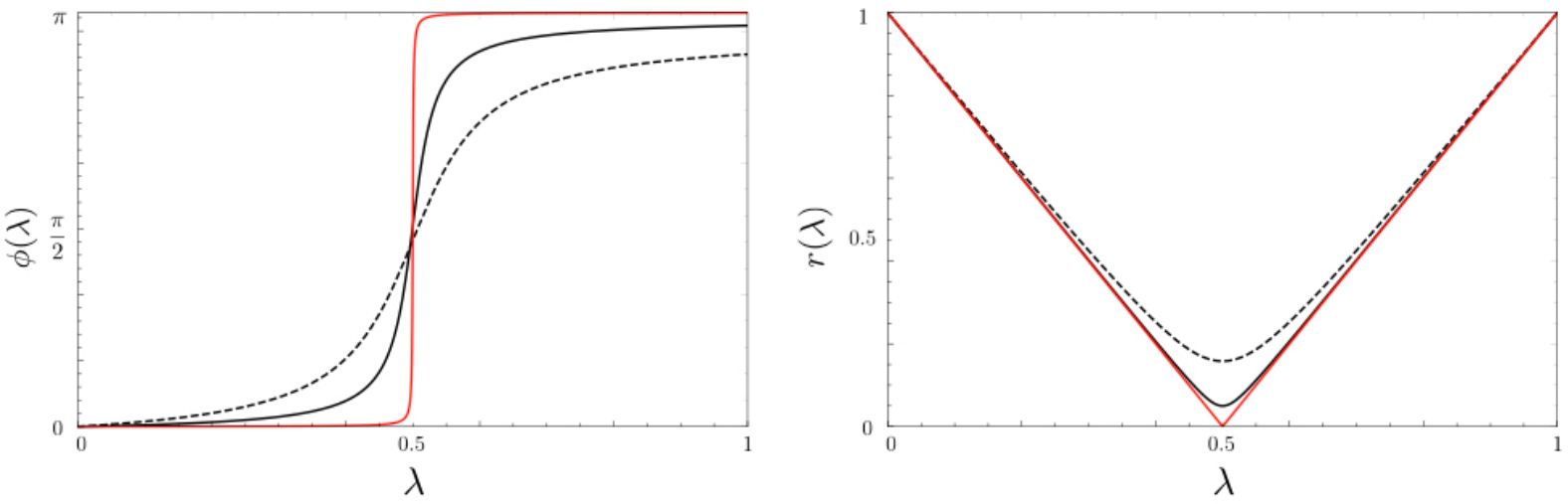}
\caption{Affinely parameterized bulk geodesic between $S$-matrix fixed points. The
dashed (solid) black line corresponds to $\delta=0.1$ ($0.01$). The red curve is approaching the discontinuous limit $\delta\to 0$.}
  \label{fig:bulkaff}
\end{figure}

In the absence of entangling forces and inaffinity, the geodesic equations in the
affine parameter $\lambda$ are:
\begin{eqnarray} 
\ddot{r} & =& \oneht r \lbrack (\dot{\phi})^2 \ +\ (\dot{\theta})^2 \rbrack \ \ , \ \ 
\ddot{\phi} \ =\ -\;2 \dot{\phi}\; {\frac{\dot{r}}{r}} \ \ , \ \
\ddot{\theta} \ =\  -\;2 \dot{\theta}\; {\frac{\dot{r}}{r}} \ .
     \label{eq:geovarrAFF}
\end{eqnarray} 
The general solution to these equations is straightforward to find. Consider the special
$S$-matrix trajectory from fixed point $\hat {\bf S}_{\tiny{\circled{1}}}$ to fixed point $\hat {\bf S}_{\tiny{\circled{2}}}$
with $\phi=\theta$. On the flat-torus, this trajectory moves along a non-entangling geodesic. In the bulk,
the corresponding geodesic is
\begin{eqnarray} 
r(\lambda ) & =& \sqrt{1+\left(4-\delta\right)\lambda\left( \lambda-1\right)} \ , \nn \\
\phi(\lambda ) & =&  \theta(\lambda ) \ =\ \tan^{-1}\!\left( \frac{\lambda \sqrt{\delta\left(4-\delta\right)}}{2-\lambda\left(4-\delta\right)} \right) \ ,
     \label{eq:geovarrAFFsol}
\end{eqnarray}
where $\lambda\in [0,1]$ and $\delta$ is a small parameter. The
boundary geodesic is recovered in the limit $\delta\to 0$ where the
trajectory goes through the singularity and there is a discontinuity
at the half-way point $\lambda=1/2$:
\begin{eqnarray} 
r(\lambda ) & =& 2|\lambda-\oneht| \ , \nn \\
\phi(\lambda ) & =&  \theta(\lambda ) \ =\ \pi\,\Theta\left(\lambda - \oneht  \right) \ ,
     \label{eq:geovarrAFFsol2}
\end{eqnarray}
where $\Theta(x)$ is the Heaviside step function. Avoidance of the singularity requires
non-vanishing $\delta$ and incurs an error in $\phi$ of order $\sqrt{\delta}$. The affine solutions are plotted for various values of $\delta$ in Fig.~(\ref{fig:bulkaff}).

\begin{figure}[!ht]
\centering
\includegraphics[width = 0.98\textwidth]{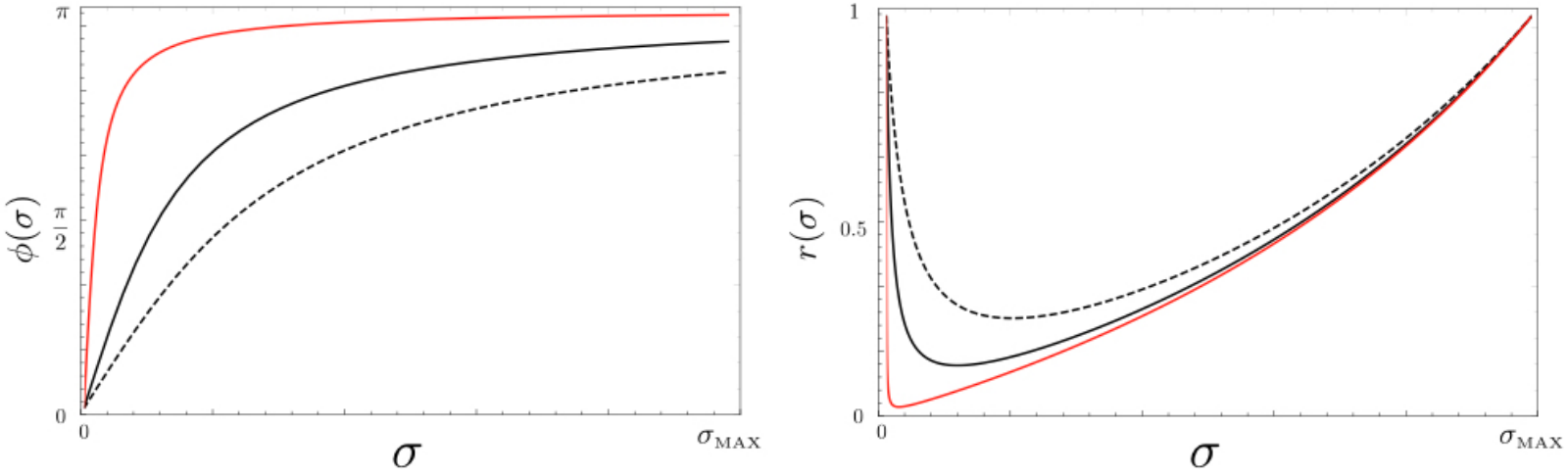}
\caption{Inaffinely parameterized bulk geodesic between $S$-matrix fixed points. The
dashed (solid) black line corresponds to $\sigmax=4$ ($8$). The red curve corresponds to $\sigmax=50$.}
  \label{fig:bulkinaff}
\end{figure}

Consider now parameterizing the same geodesic with the non-affine parameter $\sigma$.
The geodesic equations are
\begin{eqnarray} 
\ddot{r} & =& {\hat\kappa}(\sigma)\dot{r} \;+\;\oneht r \lbrack (\dot{\phi})^2 \ +\ (\dot{\theta})^2 \rbrack \ \ , \nn \\
\ddot{\phi} & =& {\hat\kappa}(\sigma)\dot{\phi}\;-\;2 \dot{\phi}\; \frac{\dot{r}}{r} \ \ , \nn \\
\ddot{\theta}   & =& {\hat\kappa}(\sigma)\dot{\theta}\; -\;2 \dot{\theta}\; \frac{\dot{r}}{r} \ \ .
     \label{eq:geovarr}
\end{eqnarray}
Choosing the inaffinity to be
\begin{eqnarray} 
  {\hat\kappa}(\sigma)  &=& \frac{\dot{\hat N}}{\hat N} \ =\ {\kappa}(\sigma) \;+\;2 \frac{\dot{r}}{r} \ ,
     \label{eq:geovarr2}
\end{eqnarray}
with $\kappa$ the inaffinity of the boundary solution given in Eq.~(\ref{eq:baseinaffinity}), then the $\phi$ and $\theta$
trajectory equations in Eq.~(\ref{eq:geovarr}) are satisfied with $\phi=\theta=-2\tan^{-1}\!\left( \sigma \right)$. However, then the first equation is solved to give
\begin{eqnarray} 
r(\sigma) & =& \frac{1+\sigma^2}{\left(1-\sigma^2\right)+ e_1 \sigma} \ ,
     \label{eq:geovarrrsol}
\end{eqnarray}
where $e_1$ is an integration constant. Note that $r$ never approaches the singularity.
While $r(0)=1$, and therefore the geodesic begins on the boundary, as $\sigma\to\infty$, $r(\sigma)\to -1$. Since $0 < r \leq 1$, the solution is sensible only up to a cutoff value of $\sigma$, $\sigmax = \frac{e_1}{2}$. Then, $\phi$ and $\theta$ only reach the fixed point up to a correction
given by $2/\sigmax$. The inaffine solutions are plotted for various values of $\sigmax$ in Fig.~(\ref{fig:bulkinaff}). By integrating the inaffinity, it is straightforward to 
relate the affine and inaffine parameterizations\footnote{One finds
\begin{eqnarray} 
\lambda \;=\; \frac{\sigma(1+\sigmax^2)}{\sigmax(1-\sigma^2 +2\sigma\sigmax)} \ ,
\end{eqnarray}
with $\delta=1/(1+\sigmax^2)$.}.
These results provide evidence that, for non-entangling geodesics,
the $S$-matrix can be recovered by going through the bulk, up to a quantifiable error.

\subsection{Recovery of the ERE and error correction}

\begin{figure}[!ht]
\centering
\includegraphics[width = 0.98\textwidth]{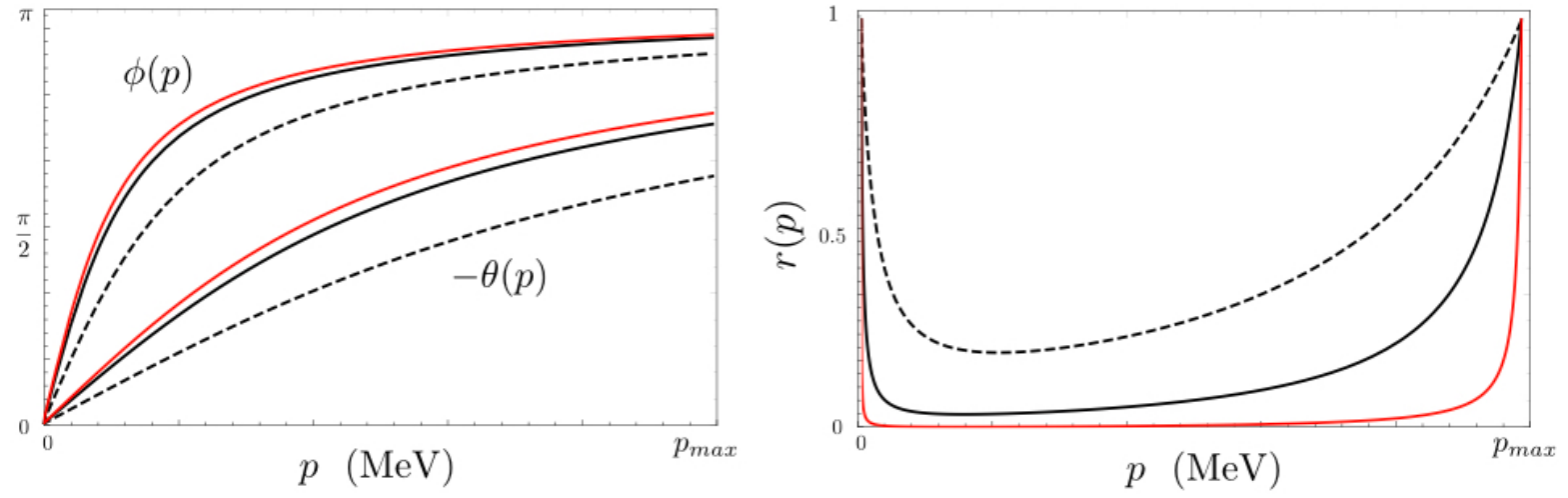}
\caption{Bulk geodesics between $S$-matrix fixed points for LO in the ERE. The
dashed (solid) black line corresponds to $\pmax=50~{\rm MeV}$ ($79~{\rm MeV}$). The red curve corresponds to $\pmax=89~{\rm MeV}$.}
  \label{fig:bulkinaffint}
\end{figure}

In the presence of an external potential, $\hat{\mathbb{V}}(r,\phi,\theta)$, with the parameter choice $\sigma=p$, the trajectory equations are 
\begin{eqnarray} 
\ddot{r} & =& {\hat\kappa}(p)\dot{r} \;+\;\oneht r \lbrack (\dot{\phi})^2 \ +\ (\dot{\theta})^2 \rbrack \;-\; \oneht{\hat N}^2 \partial_r {\hat{\mathbb{V}}} \ , \nn \\
\ddot{\phi} & =& {\hat\kappa}(p)\dot{\phi}\;-\;2 \dot{\phi}\; \frac{\dot{r}}{r} \;-\; {\hat N}^2\frac{1}{r^2} \partial_\phi {\hat{\mathbb{V}}} \ , \nn \\
\ddot{\theta}   & =& {\hat\kappa}(p)\dot{\theta}\; -\;2 \dot{\theta}\; \frac{\dot{r}}{r} \;-\; {\hat N}^2\frac{1}{r^2} \partial_\theta {\hat{\mathbb{V}}} \ .
     \label{eq:extfgeovarr}
\end{eqnarray}
Choosing the inaffinity as in Eq.~(\ref{eq:geovarr2}) but now with $N$ given by the LO ERE solution, one finds ${\hat N}=r^2{N}$. The LO ERE solution of Eq.~(\ref{eq:LOinaffineSOL}) for $\phi$ and $\theta$
is then recovered with external potential (with the choice $c_1=1$)
\begin{eqnarray}
\hat{\mathbb{V}}(r,\phi,\theta) & = & \frac{1}{r^2} \mathbb{V}(\phi,\theta) \ =\  \frac{|a_0 a_1|}{\left(|a_0|+|a_1| \right)^2}\;\frac{1}{r^2}\; \tan^2\left(\oneht(\phi+\epsilon\,\theta)\right) \ .
\label{eq:entPOTvarr} 
\end{eqnarray}
The bulk entangling potential is therefore proportional to the source of the scalar curvature of the hyperbolic space.
The differential equation for $r(p)$ is 
\begin{eqnarray} 
\ddot{r} & =& \left({\kappa}(p) +2\frac{\dot{r}}{r}\right)\dot{r} \;+\;\oneht r \big\lbrack (\dot{\phi})^2 \ +\ (\dot{\theta})^2 \ +\ 2 N^2\mathbb{V} \big\rbrack  \ .
\end{eqnarray}
The relevant solution is 
\begin{eqnarray} 
&&\hspace{-0.28in}r(p) \ =\ \cos\left( {\cal A}\oneht(\phimax-\epsilon\,\thetamax) \right)
 \sec\left( {\cal A}
  \big\lbrack (\phi(p)-\epsilon\,\theta(p))\ -\ \oneht(\phimax-\epsilon\,\thetamax)  \big\rbrack \right) \ ,
  \label{eq:nongeor}
\end{eqnarray}
where $\phimax=\phi(\pmax)$, $\thetamax=\theta(\pmax)$, and
\begin{eqnarray} 
{\cal A} \ \equiv\ \frac{\sqrt{a_0^2+a_1^2}}{\sqrt{2}(a_0-\epsilon a_1)} \ .
\end{eqnarray}
Note that $r(0)=r(\pmax)=1$. Unlike the geodesic case, there is now a maximum value of $\pmax$ beyond which the singularity is reached.
This occurs when
\begin{eqnarray} 
{\cal A}\left(\phimax-\epsilon\,\thetamax\right) &=& -\pi \ .
\end{eqnarray}
In the physical case this gives the bound $\pmax<89.4~{\rm MeV}$. Therefore, there is an intrinsic error in reproducing the physical
$S$-matrix ---that is $\phi(p)$ and $\theta(p)$--- from bulk data as the fixed point values of $\pm\pi$ can never be reached. Indeed, one finds at $\pmax$:
\begin{eqnarray} 
  \phi &=& \pi \ + \ \frac{2}{a_0 \pmax} \ + \mathcal{O}\left( (a_0 \pmax)^{-3}\right) \ =\  0.94\,\pi \ ,\nn \\
  \theta &=& -\pi \ + \ \frac{2}{a_1 \pmax} \ + \mathcal{O}\left( (a_1 \pmax)^{-3}\right) \ =\ -0.75\,\pi\ .
  \end{eqnarray}
  The bulk solutions are plotted for various values of $\pmax$ in Fig.~(\ref{fig:bulkinaffint}).
The closer the scattering length is to unitarity, the smaller the error. Conversely, natural values of the scattering lengths
incur significant errors. For the solution of the trajectory equations that is in question, $\phi$ and $\theta$ maintain the
conformal symmetry of the boundary solution, and therefore as long as the bulk data can reproduce the trajectory up to the
fixed point of the conformal transformation, $p=1/\sqrt{|a_0 a_1|}=17.4~{\rm MeV}$, the final segment of the curve can be
obtained from the reflection isometry implied by the conformal invariance.

Thus, the potential in Eq.~(\ref{eq:entPOTvarr}) has been engineered to provide a holographic dual of the nucleon-nucleon scattering matrix and,
consequently, a kind of error-correcting code, where here ``error'' is in reference to violations of unitarity due to inelastic lossiness. The existence of $\pmax$ indicates that the elastic trajectory cannot be holographically reproduced for $p > \pmax$.
Fig.~(\ref{fig:bulkdiag}) illustrates this particular error-correcting code.

\section{Projections on $\mathbb{R}^2$ and $\mathbb{R}^3$}
\label{sec:projs}

\subsection{Non-conservative entanglement forces}

\noindent The embedding of the unitarity surface in $\mathbb{R}^4$
gave rise, in the chosen coordinate system, to the flat torus, which
allowed for a straightforward solution of the trajectory
equations. However, the intrinsic four-dimensional nature of the space
renders visualization difficult. In the original considerations of
geometry, it was seen that a projection onto the
$(x,z)$ and $(y,w)$ planes gave rise to a compact rhombus which the
$S$-matrix trajectory is confined to by unitarity.  Moreover, as will
be seen below, the two unitarity constraints on the coordinates
$x,y,z,w$ can be used to describe a surface of constraint in three dimensions.  How then does one describe
these geometries?

Consider a subspace of the space described by the flat torus metric $g_{ab}$ with line element
\begin{eqnarray}
ds^2 \ =\ dx^2\;+\;dy^2\;+\;dz^2\;+\;dw^2 \ =\ \oneht\;\left( d\phi^2\;+\;  d\theta^2\right)\ .
   \label{eq:metricPROJ}
\end{eqnarray}
The metric ${\bar g}_{ab}$ of the subspace is related to the flat torus metric via
\begin{eqnarray}
g_{ab} \ =\ {\bar g}_{ab} \ +\ h_{ab} \ ,
   \label{eq:metricPROJsubs}
\end{eqnarray}
where $h_{ab}$ describes the remainder subspace of the flat torus which is left out of the projection.
With the assumption of conservative forces, the projected space is described by the trajectory equation
 \begin{equation}
\ddot{\mathcal{X}}^a \ +\ {}_{\bar g}\Gamma^a_{\ bc} \dot{\mathcal{X}}^b \dot{\mathcal{X}}^c \ =\ \kappa(\sigma) \dot{\mathcal{X}}^a
   \ -\ \oneht {\bf{N}}^2 {\bar g}^{ab}  \partial_b  {\mathbb{\bar V}} (\mathcal{X}) \ ,
    \label{eq:exEOMfromactPROJ}
 \end{equation}
 where ${}_{\bar g}\Gamma^a_{\ bc}$ are the Christoffel symbols for the metric of the projected space ${\bar g}_{ab}$.
 It is straightforward to see that the assumption of conservative forces must be abandoned.
 Expressing the trajectory equation on the flat torus, Eq.~(\ref{eq:exEOMfromact}), as
 \begin{equation}
\ddot{\mathcal{X}}^a \ +\ g^{ad}{}_{g}\Gamma_{dbc} \dot{\mathcal{X}}^b \dot{\mathcal{X}}^c \ =\ \kappa(\sigma) \dot{\mathcal{X}}^a
   \ -\ \oneht {\bf{N}}^2 {g}^{ab}  \partial_b  {\mathbb{V}} (\mathcal{X}) \ ,
    \label{eq:exEOMfromactv2}
 \end{equation}
and noting that 
 \begin{equation}
{}_{g}\Gamma_{dbc} \ =\ {}_{\bar g}\Gamma_{dbc} \ +\ {}_{h}\Gamma_{dbc} \ \ \ , \ \ \ {\bar g}_{ab}g^{bc}=2 {\bar g}_a^{\ c}
 \end{equation}
 it is straightforward to recover Eq.~(\ref{eq:exEOMfromactPROJ}) but with the substitution
 \begin{equation}
\partial_a  {\mathbb{\bar V}} \ \to \ 2 {\bar g}_a^{\ b}\partial_b  {\mathbb{V}} \ -\  \frac{2}{{\bf{N}}^2}\,{}_{\bar g}\Gamma_{abc} \dot{\mathcal{X}}^b \dot{\mathcal{X}}^c \ .
    \label{eq:barVtoV}
 \end{equation}
 Therefore, in the projected space, the entanglement force found in the original flat space is rescaled by the projected metric and in addition acquires a
 non-conservative component which exactly cancels the curvature component in the projected space. These arguments formalize the observation that the flat torus
 solution fixes the trajectories in various projections of the flat torus. However, the manner in which the trajectory equations arise in the projections will vary in that
 external forces and curvature effects may interchange their roles.
 
\subsection{Embedding in $\mathbb{R}^2$}


\noindent Consider the geometry of the rhombus. Here the $S$-matrix may be viewed in $\mathbb{R}^4$ as $\mathbb{S}_R\otimes\mathbb{R}^2$
with $\mathbb{S}_R$  denoting the rhombus. With the choice
\begin{eqnarray}
ds^2 \ =\ ds_R^2\;+\;dy^2\;+\;dw^2 \ \ \ ,\ \ \ ds_R^2 \ =\ dx^2\;+\;dz^2 
   \label{eq:R2metricgen1}
\end{eqnarray}
the line element takes the form
\begin{eqnarray}
ds_R^2 \ =\ \oneht\;\sin^2\phi\,d\phi^2\;+\; \oneht\; \sin^2\theta\,d\theta^2 \ ,
   \label{eq:KRinR2}
\end{eqnarray}
which gives the metric of the projected space:
\begin{eqnarray}
\bar g  = \oneht \begin{pmatrix}
         \sin^2\phi && 0 \\
        0 && \sin^2\theta
  \end{pmatrix} \ .
  \end{eqnarray}
Therefore, the remainder space is:
\begin{eqnarray}
h  = \oneht \begin{pmatrix}
         \cos^2\phi && 0 \\
        0 && \cos^2\theta
  \end{pmatrix} \ .
  \end{eqnarray}

The geodesic equations for an $S$-matrix trajectory with $\sigma=p$ are read off from Eq.~(\ref{eq:exEOMfromactPROJ}) to be
\begin{eqnarray}
\ddot{\phi} & =&  {\kappa}(p)\dot{\phi}\; -\;   (\dot{\phi})^2 \cot\phi \ \ \ , \ \ \
\ddot{\theta} \ =\ {\kappa}(p)\dot{\theta}\; -\;  (\dot{\theta})^2\cot\theta  \ ,
     \label{eq:KRgesNA2}
\end{eqnarray}
and the solutions are readily found. Including external forces, and making the
substitution as in Eq.~(\ref{eq:barVtoV}) immediately recovers the flat torus trajectory equations which give LO in the ERE (and the conformal range model). These solutions are plotted in
the rhombi of Section~\ref{sec:smatse}.

\subsection{Quartics and the squere}

\begin{figure}[!ht]
\centering
\includegraphics[width = 0.98\textwidth]{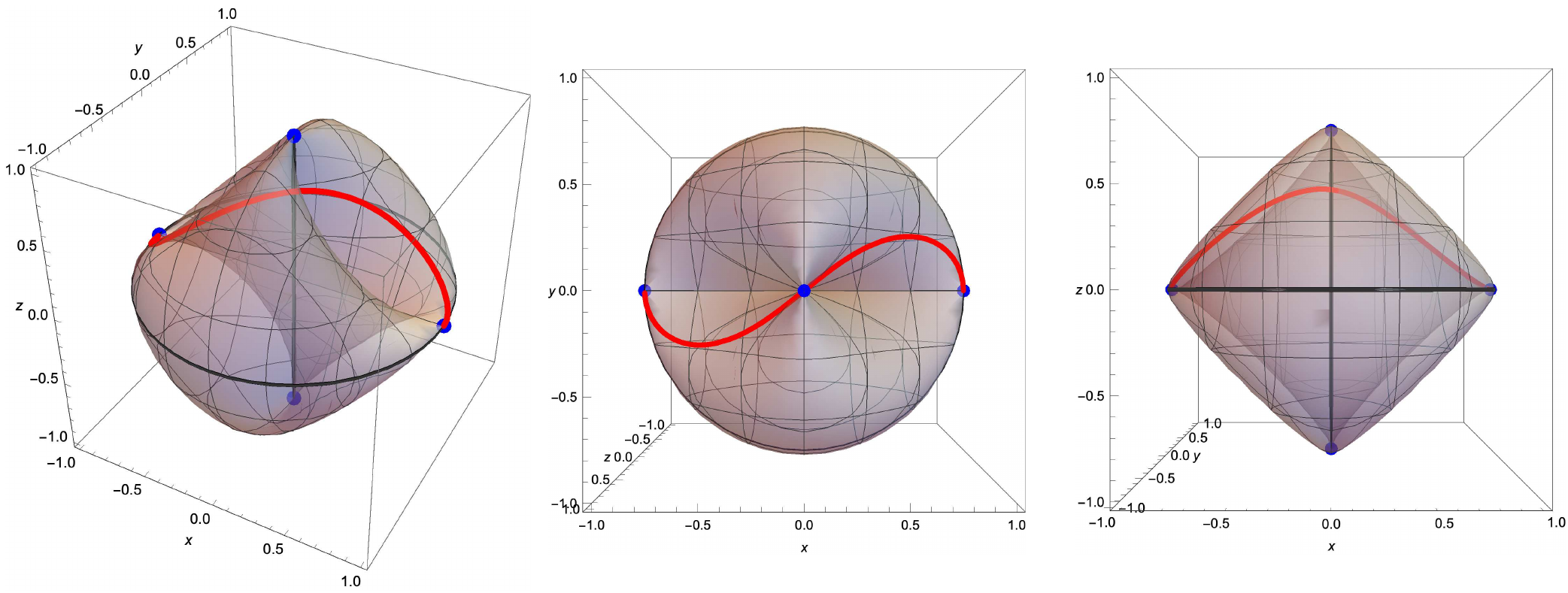}
\caption{The squere in the $x$-$y$-$z$ projection with several viewpoints of the real part of the $S$-matrix taken from LO in the ERE.
    The trajectory begins and ends at fixed points of the RG. The solid black lines are non-entangling geodesics.}
    \label{fig:squeresxyzNij}
\end{figure}

\begin{figure}[!ht]
  \centering
  \includegraphics[width = 0.98\textwidth]{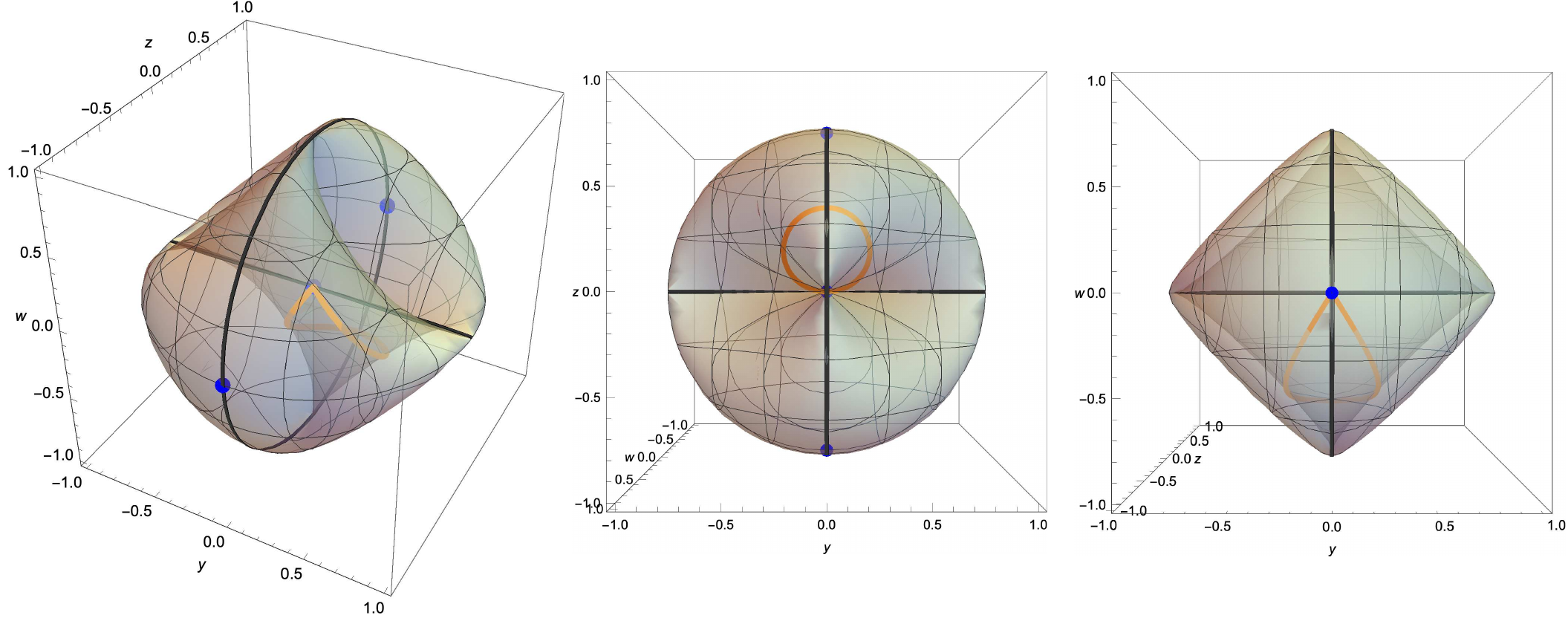}
\caption{The squere in the $y$-$z$-$w$ projection with several viewpoints of the imaginary part of the $S$-matrix taken from
  LO in the ERE.
  The trajectory begins and ends at fixed points of the RG. The solid black lines are non-entangling geodesics.}
    \label{fig:squeresyzwNij}
\end{figure}

\noindent The constraint of Eq.~(\ref{eq:S3const}) can be removed to give an equation for a quartic surface.
Eliminating the $w$ variable defines the $x$-$y$-$z$ system:
\begin{eqnarray}
y^2\left( x^2\;+\; y^2\;+\; z^2\right)\;+\; x^2z^2 \;=\; y^2 \ .
   \label{eq:NOwAC}
\end{eqnarray}
This equation describes the compact object illustrated in
Fig.~(\ref{fig:squeresxyzNij}) and  Fig.~(\ref{fig:squeresyzwNij}),
and which will be referred to as a
``squere'' because of its sphere-like and square-like properties\footnote{A parametric representation of the squere
is realized by choosing any of the three coordinates in Eq.~(\ref{eq:PS}).}.  The
unit squere fits inside the unit sphere and has volume $8/3$.  The
$S$-matrix is a trajectory that is confined to the surface of the
squere.  The fixed points of the RG, given in Eq.~(\ref{eq:FPs})
appear as four distinct points on the squere which lie on axes of
symmetry.  The regions of vanishing EP are found from
Eq.~(\ref{eq:epSnf2}) to be at $z=0$ with $x^2+y^2$ arbitrary, and at
$x=y=0$ with $z$ arbitrary, giving rise to a circle and a line, both
joining two fixed points, respectively. The fixed points and the
non-entangling regions of the squere in the $x$-$y$-$z$ system are
shown in Fig.~(\ref{fig:squeresxyzNij}).

The symmetries of the constraint equations and the positions of the
fixed points indicate that the $z$-$w$-$x$ system (elimination of $y$)
is equivalent to the $x$-$y$-$z$ system. However, eliminating the $x$
variable defines the $y$-$z$-$w$ system, which is a distinct
surface (see Fig.~(\ref{fig:squeresyzwNij}). Now two of the fixed points appear at the origin of
coordinates (and the non-entangling regions are different).
Again, the symmetries of the constraint equations and the positions of the fixed
points indicate that the $w$-$x$-$y$ system (elimination of $z$) is
equivalent to the $y$-$z$-$w$ system.

It is convenient to view the $x$-$y$-$z$ ($y$-$z$-$w$) system as the surface
on which the real (imaginary) part of the $S$-matrix lies. The nucleon-nucleon
$S$-matrix at LO in the ERE is plotted on the squere in Fig.~(\ref{fig:squeresxyzNij})
(real part) and in Fig.~(\ref{fig:squeresyzwNij}) (imaginary part). The right-most panels illustrate how the rhombus, for instance Fig.~(\ref{fig:rhomb1}), is just a particular two-dimensional cross-section of the squere.

\subsection{Embedding in $\mathbb{R}^3$}

\subsubsection*{Metric and curvature}

\noindent In order to study the geometry of the squere, the $S$-matrix may also be viewed in $\mathbb{R}^4$ as $\mathbb{S_Q}\otimes\mathbb{R}$
with $\mathbb{S_Q}$  denoting the squere. With the choice
\begin{eqnarray}
ds^2 \ =\ ds_Q^2\;+\;dw^2 \ \ \ ,\ \ \ ds_Q^2 \ =\ dx^2\;+\;dy^2\;+\;dz^2 
   \label{eq:R3metricgen1}
\end{eqnarray}
the squere metric takes the form
\begin{eqnarray}
ds_Q^2 \ =\ \onefourth\;r^2 \left(1+\sin^2\phi\right) d\phi^2\;+\; \onefourth\;r^2 \left(1+\sin^2\theta\right) d\theta^2 \;+\; \oneht \;r^2 \cos\phi\cos\theta\;d\phi\;d\theta \ .
   \label{eq:SQ1inR3}
\end{eqnarray}
The squere is an Einstein manifold with scalar curvature
\begin{eqnarray}
R \ =\ \frac{32}{r^2}\frac{\sin\phi \sin\theta}{\left(-2+ \cos2\phi+cos2\theta\right)^2} \ ,
   \label{eq:SQ1sc}
\end{eqnarray}
and vanishing Einstein tensor
\begin{eqnarray}
G_{ab} \ =\ R_{ab} \;-\; \oneht R\;g_{ab} \;=\; 0 \ .
   \label{eq:SQ1rt}
\end{eqnarray}

The geodesic equations for an $S$-matrix trajectory with $\sigma=p$ are again read off from Eq.~(\ref{eq:exEOMfromactPROJ}) to be
\begin{eqnarray}\;
\ddot{\phi}&=& \kappa(\sigma)\;\dot{\phi} -\frac{2\cos\phi}{\left(-2+ \cos2\phi+\cos2\theta\right)}
\;\big\lbrack- (\dot{\phi})^2 \sin\phi\;+\;  (\dot{\theta})^2\sin\theta\big\rbrack \ \ , \nn \\
\ddot{\theta} &=& \kappa(\sigma)\;\dot{\theta} -\frac{2\cos\theta}{\left(-2+ \cos2\phi+\cos2\theta\right)}
\;\big\lbrack  (\dot{\phi})^2\sin\phi\;-\;  (\dot{\theta})^2\sin\theta\big\rbrack  \ .
     \label{eq:SQ1gesNA}
\end{eqnarray}
and the solutions are readily found. Including external forces, and making the
substitution as in Eq.~(\ref{eq:barVtoV}) immediately recovers the flat torus trajectory equations which give LO in the ERE (and the conformal range model). Both non-entangling geodesics and LO in the ERE are plotted in Fig.~(\ref{fig:squeresxyzNij}) and Fig.~(\ref{fig:squeresyzwNij}).

\section{Summary and discussion}
\label{sec:summandconc}

\noindent The ERE is a parameterization of the $S$-matrix which
follows from the assumption that $p\cot\delta(p)$ is an analytic
function in $p^2$. It is straightforward to derive the ERE using an
EFT of contact operators which naturally gives rise to a momentum
expansion.  While the EFT is highly singular, and therefore the
operator coefficients in the EFT are renormalization scheme and scale
dependent, the ordering of operators provides a systematic expansion
and the resulting $S$-matrix is physical and matches perfectly to the
ERE.  However, while EFT distinguishes between spin-entangling and
non-spin-entangling interactions, and even reveals interesting
patterns of symmetry, it does not provide much insight into the
relative importance of these effects. Perhaps this is because the
fundamental principle underlying EFT is locality in spacetime, and
spin entanglement is intrinsically non-local.  This motivates the
search for a description of scattering that does not rely on an
expansion in local operators. Treating the $S$-matrix as the
fundamental object in the description of scattering suggests a
spacetime-independent algorithm for generating the ERE in a manner in
which entanglement is a purely geometrical property. In this paper,
such a geometrical theory of scattering has been obtained.

\vskip0.1in The main results and observations regarding the geometrical $S$-matrix theory are:

\vskip0.2in
\noindent$\bullet$ The $S$-matrix which describes low-energy
nucleon-nucleon scattering has symmetries that are not visible in the
EFT action. This observation, in itself, suggests that it is
fruitful to view the $S$-matrix as a fundamental object, in some sense
divorced from an EFT action. The symmetries of the $S$-matrix involve
inversions of the momenta, which interchange the UV and IR, and leave
linear combinations of phase shifts invariant. As such it is not
surprising that EFT is blind to these symmetries, since EFT is
by construction an IR description of the $S$-matrix.

\vskip0.2in
\noindent$\bullet$ The conformal UV/IR symmetries of the $S$-matrix
are realized as orientation reversing isometries on a compact
geometric space that is defined by unitarity. There is freedom in the
choice of coordinates that describes this space.  However, isotropy
constraints are restrictive, and a geometric embedding in
four-dimensional Euclidean space leads to a (two-dimensional) flat
torus. The $S$-matrix corresponds to a curve which lives on this
manifold and moves between RG fixed points.

\vskip0.2in
\noindent$\bullet$ $S$-matrix trajectories with vanishing entanglement
are special geodesics on the flat torus. In particular, in the absence
of entanglement, the bulk of the flat torus becomes inaccessible and $S$-matrix
trajectories evolve on a lattice with fixed points at the vertices and
special geodesics as links between the vertices. In the absence of
entanglement, the flat torus manifold does not exist.

Requiring the phase shifts to vanish at threshold uniquely picks out the non-entangling trajectory which moves along the central diagonal in Fig.~(\ref{fig:FT}). In nucleon-nucleon scattering this trajectory realizes an emergent $SU(4)$ spin-flavor symmetry of the effective action.

\vskip0.2in
\noindent$\bullet$ Non-geodesic $S$-matrix trajectories on the flat torus that entangle spins are driven by an external entangling
potential which permeates the flat torus and modifies the geodesic
equations. The resulting equations are integrable when the $S$-matrix
possesses the UV/IR conformal symmetry. This symmetry allows the explicit
construction of the potential, which is a harmonic function on the flat torus.

\vskip0.2in
\noindent$\bullet$ In contrast to EFT where the momentum expansion
emerges as a consequence of locality of the interaction, the momentum
variable in the geometric $S$-matrix formulation arises as a specific choice
of inaffine parameterization of the geodesic equations. The forces that
enter the trajectory equations are complicated non-local functions of
the momentum, as one might expect in a spacetime-independent formulation of scattering.

\vskip0.2in
\noindent$\bullet$ In the geometric formulation of scattering,
inelasticity --a non-local effect in the EFT--- is
in correspondence with the radius of a three-dimensional hyperbolic
space whose two-dimensional boundary is the flat torus. This space has
a singularity at vanishing radius, corresponding to maximal violation
of unitarity and vanishing EP. The boundary trajectory can be explicitly constructed
from a bulk trajectory with a quantifiable error, providing a simple
example of holographic duality and quantum error correction~\cite{Almheiri:2014lwa,Pastawski:2015qua}.
It may be feasible to design more intricate bulk trajectories that 
reproduce boundary properties with smaller errors.
\vskip0.2in

The holographic construction of $S$-matrix theory provides a forum for
the consideration of fundamental questions. In the AdS/CFT
correspondence~\cite{Maldacena:1997re}, the conformal field theory on
the boundary is a unitary theory and therefore the holographic
correspondence suggests that all bulk properties are consistent with
unitary evolution. In the simple hologram constructed in this
paper, the boundary space is not simply a flat space on which an
$S$-matrix propagates via unitary evolution, rather the theory space
is in some sense ``unitarity itself'' as the space is defined by
the constraints which enforce the unitarity of the $S$-matrix.
Of course there is no time and consequently no causality\footnote{In some sense, causality does constrain $S$-matrix trajectories as effective range corrections are restricted via the Wigner bound~\cite{Wigner:1955zz,Phillips:1996ae,Hammer:2010fw}.}
in the geometric $S$-matrix description and therefore implications 
for quantum gravity and spacetime in general are unclear.

A motivation for pursuing a formulation of scattering that does not
rely on spacetime locality is to obtain new insight regarding the
hierarchy of entangling operators that is observed in EFT. In this
respect, the results of this study are mixed. While the
assumption of finite and equal values of the two scattering lengths
implies Wigner's $SU(4)$ spin-flavor symmetry in the EFT action, in
the geometric construction, equal values of the scattering lengths 
constrains $S$-matrix trajectories to lie on geodesics that connect
fixed points. There is therefore enhanced discrete symmetry
on the flat torus when the scattering lengths are equal ---{\it i.e.} 
$\phi=\theta$--- and the $S$-matrix trajectory lies on a non-entangling 
geodesic.  However, on the flat torus there is also enhanced discrete
symmetry when $\phi=-\theta$ and the trajectory lies on an entangling 
trajectory. This asymmetry of entanglement reflects the entanglement power's
preferred direction on the flat torus: lines of equi-entanglement are 
always parallel to $\phi=\theta$. The flat-torus construction does not 
seem to provide any insight into the origin of this asymmetry.

A more-specialized motivation of this new formulation of scattering theory is to 
learn about mysterious aspects of the nucleon-nucleon EFT construction that emerge
as soon as the pion is included as a fundamental degree of freedom\footnote{For a recent
  review, see Ref.~\cite{vanKolck:2020llt}}. The pion interactions
give rise to a tensor force in spin-triplet channels which wreaks havoc on
the original EFT construction formulated in
Ref.~\cite{Weinberg:1990rz,Weinberg:1991um}. For instance, in the
chiral limit, the $\siii$ interaction is governed by an attractive
$1/r^3$ potential whose singular nature has confounded attempts to
provide a compelling regularization and renormalization scheme which
would allow a description of nucleon-nucleon interactions within a
single, unified, and systematic
framework~\cite{Kaplan:1996xu,Fleming:1999ee,Beane:2001bc,Nogga:2005hy,PhysRevC.74.014003,Kaplan:2019znu}. Here
the focus has been on understanding the geometric $S$-matrix theory of
delta-function interactions. A natural generalization would be to
consider non-relativistic systems with other varieties of singular
potentials. It may be that the geometric construction, which does not
make an {\it a priori} separation between short- and long-distance
effects, provides insight regarding the renormalization of the
nucleon-nucleon EFT relevant at energies where the pion should be included
as a fundamental degree of freedom.

One may wonder about the generality of the geometric $S$-matrix
construction, and in particular whether some general underlying
conditions that give rise to such a description can be
specified. Given the emphasis on the low-energy nucleon-nucleon system
and the expression of the $S$-matrix as four-by-four matrices in the
outer-product spin space of the two scattering nucleons, it may seem
that this construction relies in some way on spin.  However, even in
the nucleon-nucleon system the construction does not rely on spin;
{\it a priori} the nucleon-nucleon system appears symmetric under 
interchange of spin and isospin and it is only via the choice of spin 
basis that one speaks of spin entanglement. Presumably a choice of basis which 
emphasizes isospin  and of course leaves the $S$-matrix unchanged would 
lead instead to isospin entanglement\footnote{In the EFT, this is made manifest by
  Fierz transformation of the four-nucleon operators.}. A simple
condition which evidently should be satisfied in order to obtain a
useful geometric formulation is that the codimension of the
$S$-matrix trajectory in the space of all possible trajectories
consistent with unitarity must be greater than or equal to one. So,
for instance, the single-channel case with codimension-zero has no
useful geometric description, whereas the nucleon-nucleon s-wave with
codimension-one has a rich geometric representation. Other scattering
systems and interactions with more than two bodies may provide further
insight regarding this question.

\section*{Acknowledgments}

\noindent We would like to thank Zohreh Davoudi, David B.~Kaplan,
Natalie Klco, Martin J.~Savage, and Hersh Singh for important
discussions. This work was supported by the U.~S.~Department of Energy
grants {\bf DE-FG02-97ER-41014} (UW Nuclear Theory) and {\bf DE-SC0020970}
(InQubator for Quantum Simulation).

\bibliographystyle{JHEP}
\bibliography{bibi}

\end{document}